\documentclass[a4paper,11pt]{article}
\usepackage{jheppub}
\usepackage[T1]{fontenc}
\usepackage{epsfig}
\usepackage{latexsym}
\usepackage{graphicx}
\usepackage{amsmath}
\usepackage{amsfonts}   
\usepackage{amssymb}    
\usepackage{float}
\usepackage{bm}
\usepackage{url}
\usepackage{hyperref} 
\usepackage[nodisplayskipstretch]{setspace}
\def\met{\ensuremath{E_{\mathrm{T}}^{\mathrm{miss}}}}

\def\lsim{\lesssim}
\def\gsim{\gtrsim}

\def    \beq            {\begin{equation}}
\def    \eeq            {\end{equation}}
\def    \bea           {\begin{eqnarray}}
\def    \eea           {\end{eqnarray}}

\def \mn{\mu\nu{\rm SSM}}

\def\g2{{\rm GeV}^2}

\def\sw2{sin^2 \theta_w}

\def\a^tau{\alpha_{\tau}}

\def\beq{\begin{equation}}
\def\eeq{\end{equation}}
\def\beqa{\begin{eqnarray}}
\def\eeqa{\end{eqnarray}}

\newcommand{\newc}{\newcommand}
\newc\BR{BR}
\newc{\akappa}{A_{\kappa} }
\newc\deltagmtwo{\delta (g-2)_{\mu}} 
\newc\deltaamu{\Delta a_{\mu}}

\def\anti{\overline}

\def\la{\lambda}

\def\rpv{{R}_{p} \hspace{-0.4cm}\slash\hspace{0.2cm}}

\newc{\haa}{BR\(h_1\to a_1 a_1\)}
\newc{\abb}{BR\(a_1\to b\anti{b}\)}
\newc{\hbb}{BR\(h_1\to b\anti{b}\)}
\newc{\abund}{\Omega h^2}
\newc\bsgamma{b\rightarrow s \gamma }
\newc\bxsgamma{\overline{B}\rightarrow X_{s}\gamma}
\newc\brbsgamma{\BR(\overline{B}\rightarrow X_s\gamma)}

\allowdisplaybreaks
\usepackage{xcolor}
\usepackage{booktabs}
\usepackage{subcaption} 
\usepackage{colortbl}

\definecolor{LightGray}{cmyk}{0,0,0,0.2}

 \title{
Looking for the left sneutrino LSP with displaced-vertex searches
}

\author[a,b]{I\~naki Lara,}
\author[c,d]{Daniel~E.~L\'opez-Fogliani,}
\author[a,b]{Carlos~Mu\~noz,} 
\author [e]{Natsumi Nagata}
\author [f]{Hidetoshi Otono}
\author[g]{and Roberto~Ruiz~de~Austri} 
    \affiliation[a]{Departamento de F\'{\i}sica Te\'{o}rica, Universidad Aut\'{o}noma de Madrid (UAM),
Campus de Cantoblanco, 28049 Madrid, Spain}
\affiliation[b]{Instituto de F\'{\i}sica Te\'{o}rica UAM-CSIC, 
  Campus de Cantoblanco, 28049 Madrid, Spain}
  \affiliation[c]{Instituto de F\'isica de Buenos Aires UBA \& CONICET, Departamento de F\'isica,
 Facultad de Ciencia Exactas y Naturales, Universidad de Buenos Aires, 
1428 Buenos Aires, Argentina}
\affiliation[d]{Pontificia Universidad Cat\'olica Argentina, 
1107 Buenos Aires, Argentina}
\affiliation[e] {Department of Physics, University of Tokyo, 
Tokyo 113-0033, Japan}
\affiliation[f] {Research Center for Advanced Particle Physics, Kyushu University, 
Fukuoka 819-0395, Japan}
  \affiliation[g]{Instituto de F\'{\i}sica Corpuscular CSIC--UV, 
c/ Catedr\'atico Jos\'e Beltr\'an 2, 46980 Paterna,
 Valencia, Spain}
\emailAdd{inaki.lara@csic.es}
\emailAdd{daniel.lopez@df.uba.ar}
\emailAdd{c.munoz@uam.es}
\emailAdd{natsumi@hep-th.phys.s.u-tokyo.ac.jp}
\emailAdd{otono@phys.kyushu-u.ac.jp}
\emailAdd{rruiz@ific.uv.es}

\abstract{ 
We analyze a displaced dilepton signal expected at the LHC for a tau
left sneutrino as the lightest supersymmetric particle with a mass in
the range $45$--$100$~GeV. The sneutrinos are pair produced via a virtual $W$, $Z$ or $\gamma$ in the $s$ channel and,
given the large value of the tau Yukawa coupling, their decays into
two dileptons or a dilepton plus missing transverse energy from
neutrinos can be significant. The discussion is carried out in the framework of the $\mn$,
where the presence of $R$-parity violating couplings involving
right-handed neutrinos solves the $\mu$ problem and can reproduce the
neutrino data. 
{To probe the tau left sneutrinos we
compare the predictions of this scenario
with the ATLAS search for
long-lived particles using displaced lepton pairs in $pp$ collisions at
$\sqrt s= 8$~TeV, allowing
us to constrain the parameter space of the model.
We also consider an optimization of the
trigger requirements used in existing displaced-vertex searches by means
of a High Level Trigger that exploits tracker information. This
optimization is generically useful for a light metastable particle
decaying into soft charged leptons. 
The constraints on the sneutrino turn out to be more stringent.
We finally discuss the
prospects for the 13~TeV LHC searches as well as further potential
optimizations.} 
}

\preprint{\begin{flushright}
          \hspace*{3cm} 
IFT-UAM/CSIC-18-026
 \\
FTUAM-18-8
\\
UT-18-05
\\
KYUSHU-RCAPP-2018-002

            \end{flushright}}

\date{\today}

\begin{document}
\maketitle

\section{Introduction}







Clarify whether or not low-energy supersymmetry (SUSY) exists in Nature
is one of the main goals of the LHC. If $R$-parity ($R_p$) is not conserved, the lightest supersymmetric particle (LSP) decays into standard model particles 
and displaced vertices can appear, implying that dedicated analyses to detect
SUSY are necessary.
The 
`$\mu$ from $\nu$'
supersymmetric standard model ($\mn$)~\cite{LopezFogliani:2005yw,Escudero:2008jg}, is a SUSY framework
where the $\mu$ problem is solved and the neutrino data can be reproduced~\cite{LopezFogliani:2005yw,Escudero:2008jg,Ghosh:2008yh,Bartl:2009an,Fidalgo:2009dm,Ghosh:2010zi}. For this mechanism to work, the presence of trilinear
couplings involving right-handed neutrino superfields is crucial. 
As a consequence, $R_p$ is explicitly broken ($\rpv$).
Given the
$\rpv$
in the $\mn$, essentially all SUSY particles are potential candidates for LSPs, and therefore dedicated analyses of the LHC phenomenology associated to each candidate are crucial to test SUSY.
As we will discuss below, in this work we will focus our analysis on the left sneutrino as the LSP candidate.

The simplest superpotential of the $\mn$~\cite{LopezFogliani:2005yw,Escudero:2008jg} is built with one right-handed neutrino superfield $\hat\nu^c$:
\bea
W = &&
\epsilon_{ab} \left(
Y_{e_{ij}}
\, \hat H_d^a\, \hat L^b_i \, \hat e_j^c +
Y_{d_{ij}} 
\, 
\hat H_d^a\, \hat Q^{b}_{i} \, \hat d_{j}^{c} 
+
Y_{u_{ij}} 
\, 
\hat H_u^b\, \hat Q^{a}
\, \hat u_{j}^{c}
\right)
\nonumber\\
&+&   
\epsilon_{ab} \left(
Y_{{\nu}_{i}} 
\, \hat H_u^b\, \hat L^a_i \, \hat \nu^c 
-
\lambda \, \hat \nu^c\, \hat H_u^b \hat H_d^a
\right)
+
\frac{1}{3}
\kappa
\hat \nu^c\hat \nu^c\hat \nu^c,
\label{superpotential}
\eea
where the summation convention is implied on repeated indexes, with  
$a,b=1,2$ $SU(2)_L$ indices
and $i,j,k=1,2,3$ the usual family indices of the standard model. In this equation
we have neglected the conventional trilinear $\rpv$ couplings (see e.g. the discussion 
in Ref.~\cite{Lopez-Fogliani:2017qzj}). 
The terms 
in the second line 
are characteristic of the $\mn$.
The first one 
contains the Dirac Yukawa couplings for neutrinos.
The last two, after the electroweak symmetry breaking induced by the
soft SUSY-breaking terms,
generate dynamically
the $\mu$ term, 
$\mu=\la \frac{v_{R}}{\sqrt 2}$,
and the Majorana mass for the right-handed neutrino $\nu_R$, 
$m_{\mathcal M}={2}\kappa \frac{v_{R}}{\sqrt 2}$, 
respectively. With the choice of CP conservation, we have defined
the vacuum expectation value (VEV) of the right sneutrino 
$\widetilde \nu_{R}$, as 
$\langle \widetilde \nu_{R}\rangle = \frac{v_{R}}{\sqrt 2}$. 
SUSY particles do not appear in pairs in these terms, thus
they generate $\rpv$ couplings
harmless with respect to proton decay.
In the limit $Y_{\nu_i} \to 0$, $\hat \nu^c$ can be identified in the superpotential 
as a
pure singlet superfield without lepton number, and $R_p$ is not broken.
Thus, the neutrino Yukawa couplings $Y_{\nu_{i}}$ are the parameters which determine the $\rpv$ in the $\mn$, and as a consequence
this violation is small.

In addition to the VEV for the $\widetilde \nu_{R}$,
the minimization of the scalar potential induces VEVs for the neutral 
Higgses $H_{d}^0$ and 
$H_{u}^0$, $\langle H_{d,u}^0\rangle = \frac{v_{d,u}}{\sqrt 2}$, and
the left sneutrinos $\widetilde \nu_i$,
$\langle \widetilde \nu_{i}\rangle = \frac{v_{i}}{\sqrt 2}$,
with {$v_{i}\sim Y_{\nu_i} v_u$}.
The new couplings and sneutrino VEVs in the $\mn$ induce new mixing of states.
The associated mass matrices were studied in detail in
Refs.~\cite{Escudero:2008jg,Bartl:2009an,Ghosh:2017yeh}.
Summarizing, {in the case of one right-handed neutrino superfield}, there are 
eight neutral fermions (neutralinos-neutrinos),
five charged fermions (charginos-leptons), 
seven charged scalars (charged Higgses-sleptons),
and
six neutral scalars and five neutral pseudoscalars (Higgses-sneutrinos).
It is worth noticing here that whereas $v_R\sim$ TeV,
$v_{i}\lsim 10^{-4}$ GeV because of the small contributions 
$Y_{\nu_i} \lsim 10^{-6}$
whose size is determined by the electroweak-scale 
seesaw of the $\mn$~\cite{LopezFogliani:2005yw, Escudero:2008jg}.
In fact, this is a generalized seesaw since 
the left-handed neutrinos mix not only with the right-handed one, but also with the neutralinos. Given that the mass matrix is of rank 6, only
one of the light neutrinos gets a non-vanishing tree-level contribution to its mass, whereas the other two get their masses at one loop. The tree-level mass is given by:
\bea
m_{\nu} =\frac{1}{4 M_{\text{eff}}}\sum_i
\left[v_{i}^2
+ v_d \left(
\frac{2 v_{i} Y_{{\nu}_i}}{\lambda}  
+\frac{v_d Y^2_{{\nu}_i}}{\lambda^2}
\right)\right],
\label{neumass}
\eea
with
\bea
M_{\text{eff}} \equiv M\left[1 -  \frac{v^2}{ \sqrt 2 M(\kappa \, v^2_R  + \lambda v_{d} v_{u}) \, \lambda v_R }
\left(\kappa \,  v^2_R  \frac{v_d v_u}{v^2} + \frac{1}{4}\lambda v^2\right)\right],
\label{effmass}
\eea
where $v^2 \equiv v_d^2 + v_u^2$, and 
\begin{equation}
\frac{1}{M}\equiv\frac{g'^2}{M_1} + \frac{g^2}{M_2}~.
  \label{eq:3.550}
\end{equation}
Notice that
we can simplify Eq.~(\ref{neumass}) taking into account that
the second term is proportional to $v_d$, and therefore considering it
negligible in the limit of large or even moderate $\tan\beta \equiv v_u/v_d$ provided that 
$\lambda$ is not too small. 
In Eq.~(\ref{effmass}), the second term is also negligible in this limit, and for typical values of the parameters involved in the seesaw also the third one, i.e.
$M_{\text{eff}}\approx M$.
Under these assumptions, the first
term in Eq.~(\ref{neumass}) is generated only through the mixing of left-handed neutrinos with gauginos, and we arrive to the approximate formula
$m_{\nu} \approx {\sum_i {v_{i}^2}}/{4M}$.

A promptly decaying sneutrino as the LSP was analyzed in Ref.~\cite{Ghosh:2017yeh}, with a decay length $\lsim$ 0.1 mm. In this work, we will analyze
the interesting case of displaced vertices of the order of the millimeter, when a sneutrino 
in the range of masses $45-100$ GeV is allowed. The lower bound imposed not to disturb the 
decay width of the $Z$.
We will focus on the simplest case of the $\mn$ with one
right-handed neutrino superfield discussed above, and leave
the case of three families 
where all the neutrinos get contributions to their masses at tree level for a forthcoming publication~\cite{prepa}.

In the $\mn$, the mixing between left sneutrinos and Higgses/right sneutrino is small, and
this implies that the 
left sneutrino
dominates the LSP composition. 
Besides, scalar and pseudoscalar states,
{which in our convention are defined as 
$\widetilde{\nu}_{i} = 
\frac{1}{\sqrt 2} 
(\widetilde{\nu}_{i}^\mathcal{R} 
+i \widetilde{\nu}_{i}^\mathcal{I})$ , 
are co-LSPs since they have essentially degenerate masses,
$m_{\widetilde{\nu}^{\mathcal{R}}_{i}}
\approx
 m_{\widetilde{\nu}^{\mathcal{I}}_{i}}
\equiv 
m_{\widetilde{\nu}_{i}}$. Their approximate tree-level value is
\bea
m_{\widetilde{\nu}_{i}}^2
\approx  
\frac{Y_{{\nu}_i}v_u}{2v_{i}}v_{R}
\left(
-\sqrt 2 A_{{\nu}_i}-\kappa v_{R}
+
\frac{\lambda v_{R}}{\tan\beta}
\right),
\label{evenLLL2}
\eea
where $A_{{\nu}_i}$ are the trilinear parameters in the soft Lagrangian, $-\epsilon_{ab} T_{{\nu}_{i}} H_u^b \widetilde L^a_{iL} \widetilde \nu_{R}^*$,
defining $T_{{\nu}_{i}}\equiv A_{{\nu}_i} Y_{{\nu}_i}$ and with
the summation convention not applied in this case. 
Taking all this into account, scalar and pseudoscalar sneutrinos 
are dominantly pair-produced via a Drell-Yan process mediated by a virtual $W$, $Z$ or $\gamma$.
Besides, the case of the tau left sneutrino 
($\widetilde{\nu}_{\tau}$)
LSP turns out to be particularly interesting, because of the large value of the tau Yukawa coupling which can give rise to significant branching ratios (BRs) for decays to\footnote{In what follows, the symbol $\ell$ will be used for an electron or a muon,
{$\ell=e,\mu$,}
and charge conjugation of fermions is to be understood where appropriate.}
$\tau\tau$ and $\tau \ell$.


To probe the $\widetilde{\nu}_{\tau}$
LSP, the dilepton displaced-vertex
searches are found to be most promising. 
{We compare the 
predictions of our scenario
with the ATLAS search~\cite{Aad:2015rba} for long-lived
particles using displaced lepton pairs $\ell \ell$ in $pp$ collisions at
$\sqrt s= 8$~TeV, which
allows us to constrain the parameter space of the model.
Nevertheless, the existing searches
\cite{Aad:2015rba, CMS:2014hka} are designed for a generic purpose and
thus not optimized for light metastable particles such as 
the $\widetilde{\nu}_{\tau}$.
We therefore consider also a possibility of improving these searches
by lowering trigger thresholds, relying on a High Level Trigger that
utilizes tracker information. As it turns out, this optimization is
quite feasible and considerably improves the sensitivity of the 
displaced-vertex searches. 
We also
consider an optimization of the 13~TeV LHC searches and show the
prospects for investigating the 
parameter space of our scenario by searching for the
$\widetilde{\nu}_{\tau}$
at the 13~TeV LHC run. Possibilities of further
improvements for these searches will also be discussed.}

The paper is organized as follows. In Section~\ref{phenomenology}, we will introduce
the phenomenology of the 
$\widetilde{\nu}_{\tau}$ 
LSP,
studying its pair production channels at the LHC, as well as the signals.
These consist of two dileptons or a dilepton plus missing transverse energy (MET) 
from the 
sneutrino
decays. On the way, we will analyze the decay widths, BRs and decay lengths.
{In Section~\ref{reinterpretation}, we consider first
the
existing dilepton displaced-vertex searches, and discuss
its feasibility and significance on $\widetilde{\nu}_{\tau}$
searches.
Then, we study 
an optimization by using a High Level
Trigger with tracker information.}
We also show our
prescription for recasting the ATLAS 8-TeV result \cite{Aad:2015rba} to
the case of the
$\widetilde{\nu}_{\tau}$.
We then show the current reach of this
search on the 
parameter space of our scenario based on
the ATLAS 8-TeV result \cite{Aad:2015rba}, and the prospects for the
13-TeV searches in Section~\ref{results}. Our conclusions and prospects
for future work are presented in Section~\ref{conclusions}.

\section{Left sneutrino LSP phenomenology in the $\mn$}

\label{phenomenology}

\begin{figure}[t!]
\centering
  \subcaptionbox{
\label{fig:production31a} $Z$ channel}{\includegraphics[scale=0.45]{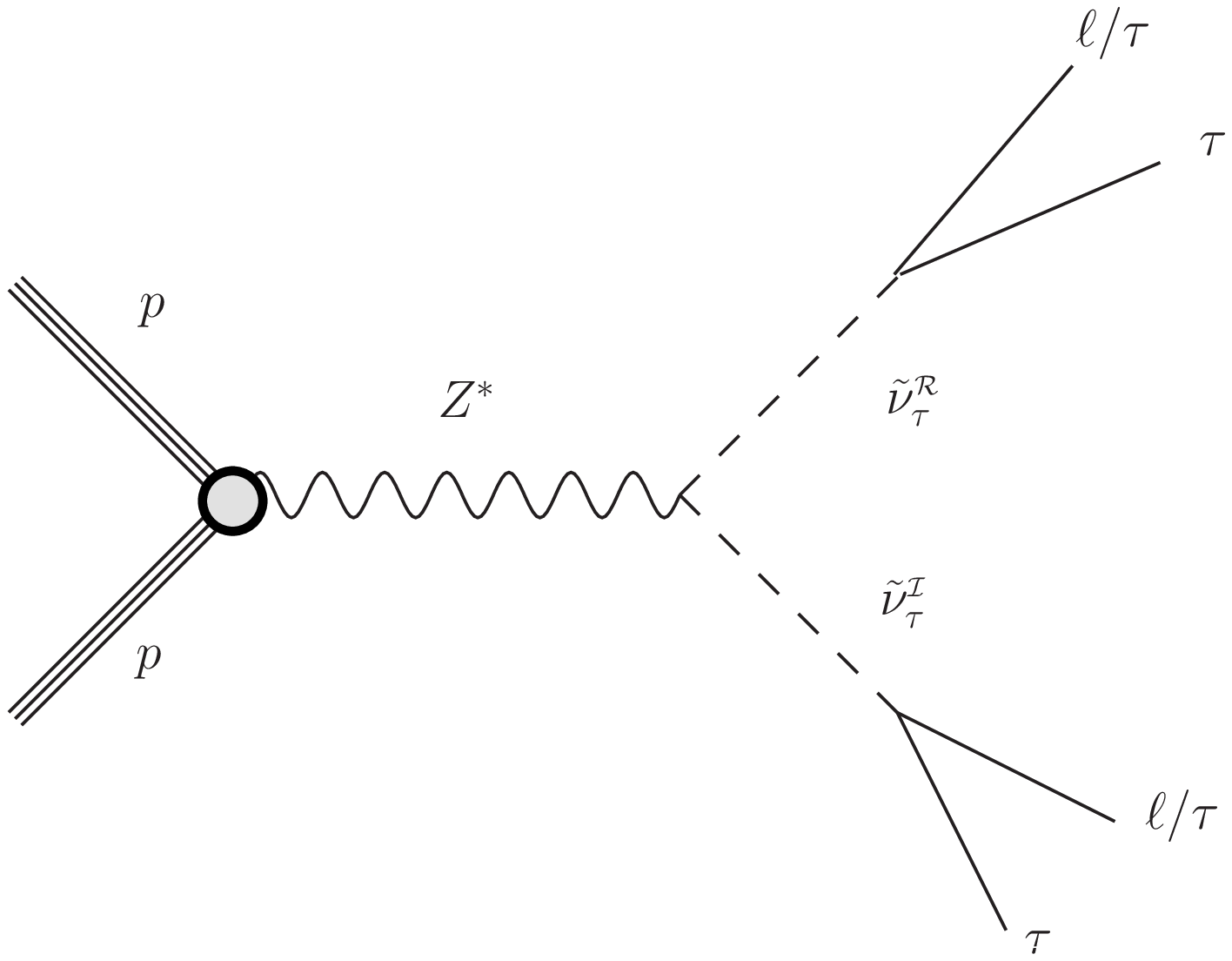}} 
  \subcaptionbox{
\label{fig:production31c} $\gamma ,Z$ channels}{\includegraphics[scale=0.45]{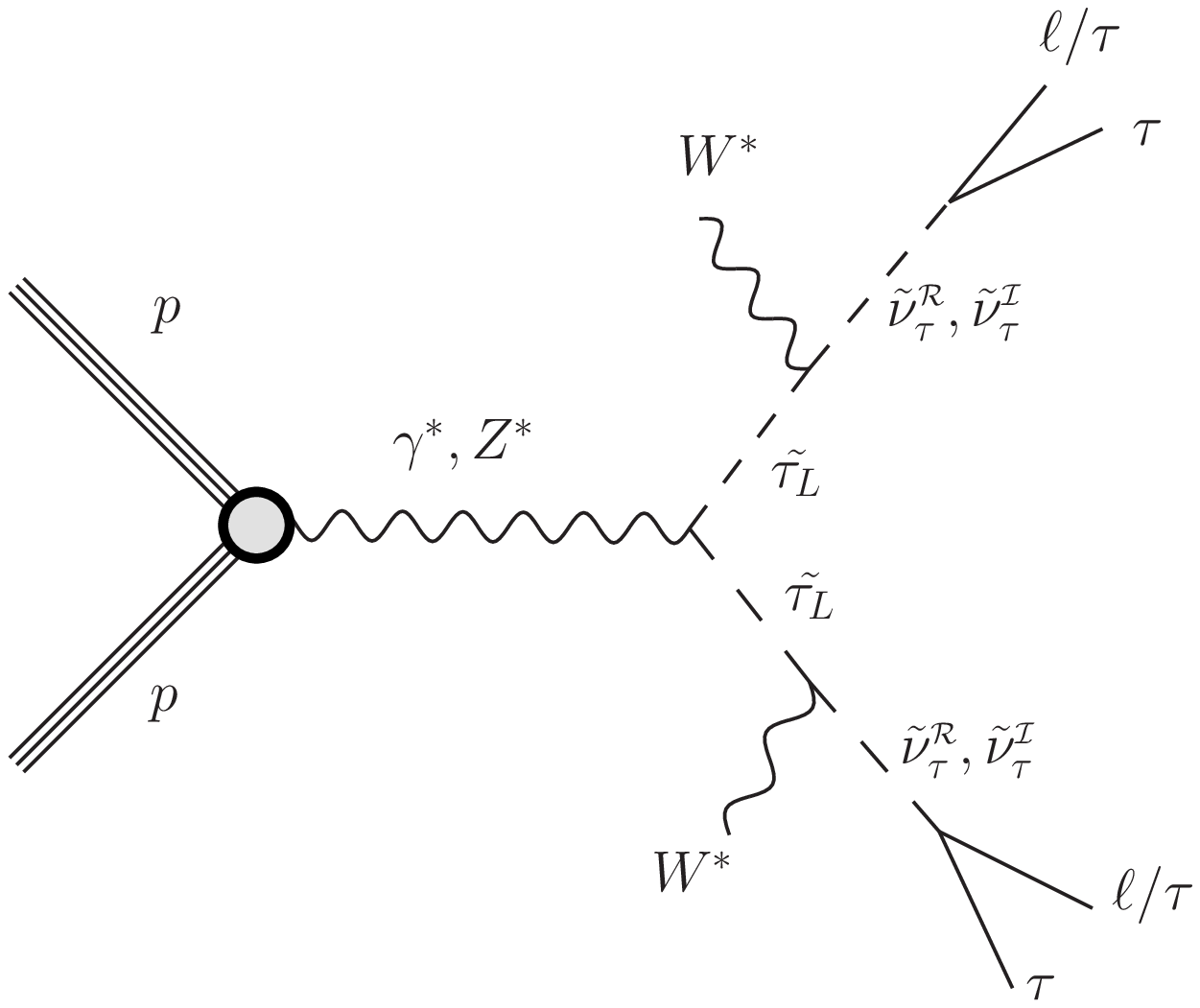}}
  \subcaptionbox{
\label{fig:production31b} $W$ channel}{\includegraphics[scale=0.45]{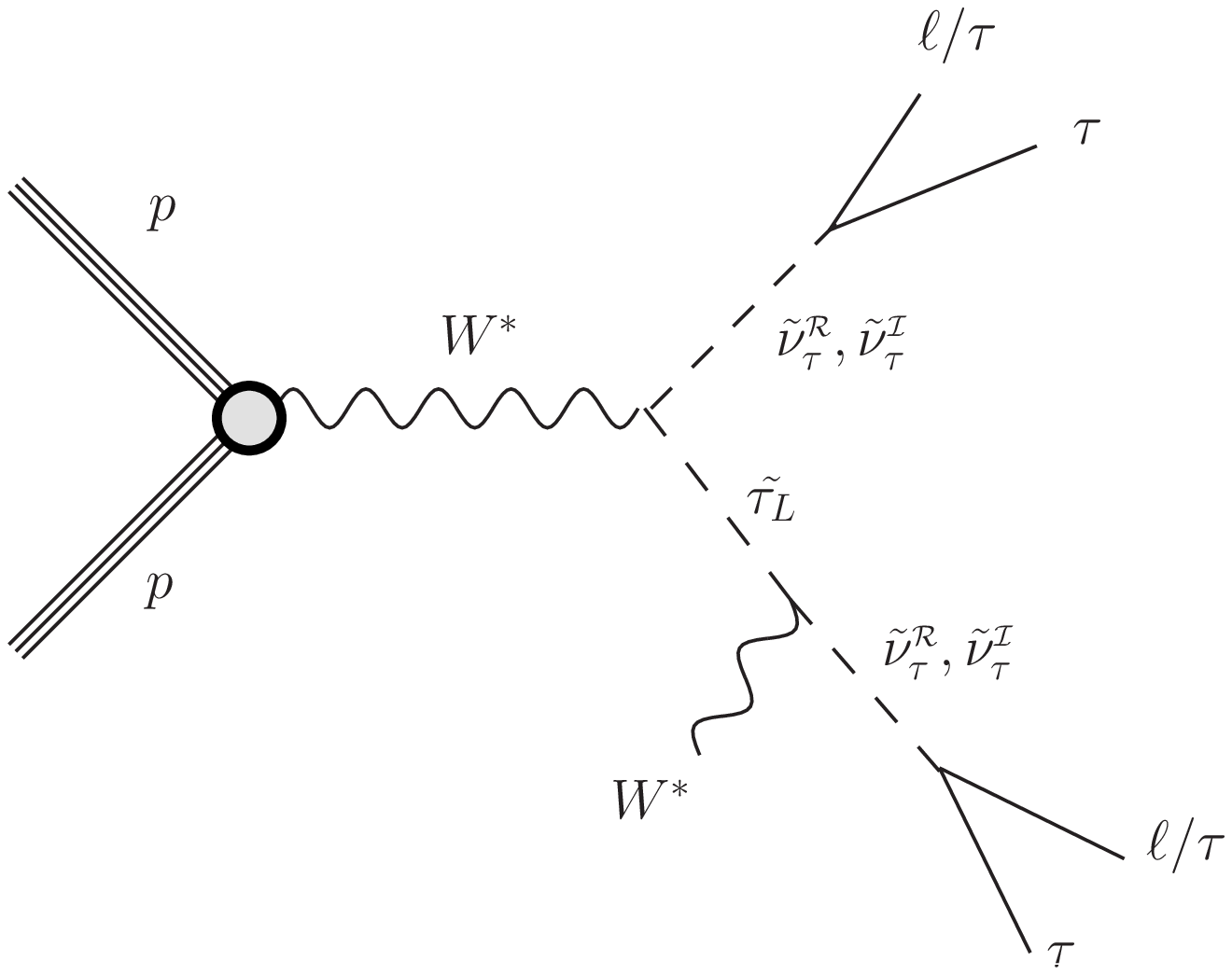}}
\caption{Decay channels into two 
$\tau\,\ell/\tau$,
from a pair production at the LHC
of scalar and pseudoscalar tau left sneutrinos co-LSPs.
{Decay channels into one $\tau\,\ell/\tau$ plus neutrinos are the same but substituting 
in (a), (b) and (c) one of the two vertices by a two-neutrino vertex}.
}
\label{fig:production31}
\end{figure} 

{The dominant pair production 
channels of sleptons at large hadron colliders were studied 
in 
Refs.~\cite{Dawson:1983fw,Eichten:1984eu,delAguila:1990yw,Baer:1993ew,Baer:1997nh,Bozzi:2004qq}.
In Fig.~\ref{fig:production31}, we can see the production 
channels at the LHC for the case of the $\widetilde{\nu}_{\tau}$ LSP which is the one interesting for our analysis.}
The direct production of Fig.~\ref{fig:production31a}
occurs via a $Z$ channel giving rise to a pair of scalar and pseudoscalar left sneutrinos. 
As discussed in the introduction,
these states
have 
essentially degenerate masses and are therefore co-LSPs.
On the other hand, since the left stau is
in the same $SU(2)$ doublet as the tau left sneutrino, it becomes the
next-to-LSP (NLSP).
{The mass splitting is mainly due to the usual D-term contribution. Similar to the MSSM, one obtains that the difference of the squared masses is given by $-m_W^2 \cos 2\beta$. Taking into account the particular values of $\tan\beta$ and the sneutrino mass (as well as the loop corrections to sneutrino and stau masses), the typical mass difference is of about 20--30 GeV}.
Thus the direct production and decay of the left stau is another important source of the $\widetilde{\nu}_{\tau}$ LSP.
In particular, pair production can be obtained
through a $\gamma$ or $Z$ decaying into two staus, as shown in Fig.~\ref{fig:production31c},
with the latter
dominantly decaying into 
a (scalar or pseudoscalar) sneutrino plus an off-shell $W$ producing a
soft meson or a pair of a charged lepton and a neutrino.
Besides, sneutrinos can be pair produced through a 
$W$ decaying into a stau and a (scalar or pseudoscalar) sneutrino
as shown in Fig.~\ref{fig:production31b}, with the stau 
decaying as before.
{The number of sneutrino pairs produced through these channels at 8 TeV and 13 TeV for integrated luminosities of 20.3 fb$^{-1}$ and 300 fb$^{-1}$, respectively, are shown in Table~\ref{sneutrino_Xsection}.}



\begin{table}[t]
 \caption{Number of sneutrino pairs produced through the different channels shown in Fig.~\ref{fig:production31} at 8 TeV and 13 TeV  with integrated luminosity of 20.3 fb$^{-1}$ and 300 fb$^{-1}$, respectively.}
\label{sneutrino_Xsection}
{\footnotesize
\begin{minipage}[b]{0.45\linewidth}\centering
  \begin{tabular}{c|*4c|*4c}
 \toprule
&\multicolumn{4}{c}{8 TeV }&\multicolumn{4}{c}{13 TeV }\\
 \midrule
  $m_{\widetilde{\nu}_{\tau}}$ (GeV)&50&60&80&100&50&60&80&100\\  \hline
   Through $Z$ boson&66,400&22,370&6,090&2,448&1,794,000&623,000&179,800&75,700\\ 
   Through $W$ boson&   66,600&34,870&12,840&5,915& 1,860,000&997,000&385,600&184,800\\  
   Through $\gamma ,Z$ bosons&6,585&3,954&1,703&851&226,300&139,400&62,550&32,683\\   
      \bottomrule
     \end{tabular}
 \end{minipage}
 }
\end{table}

In Fig.~\ref{fig:production31}, we also show the detectable decay
of the pair-produced $\widetilde{\nu}_{\tau}$
into $\tau\,\ell/\tau$.
As a result of the mixing between left sneutrinos and Higgses, 
the sizable decay of $\widetilde{\nu}_{\tau}$ into $\tau\tau$ is possible
because of the large value of the tau Yukawa coupling.  
Other sizable decays into $\tau\,\ell/\tau$ can occur through the Yukawa interaction of 
$\widetilde{\nu}_{\tau}$ with $\tau$ and charged Higgsinos,
via the mixing between the latter and $\ell$ or $\tau$.
To analyze these processes we can write approximate formulas for
the partial decay widths of the scalar/pseudoscalar tau left sneutrino.
The one
into 
$\tau\tau$ is given by: 
%
\begin{equation}
  \Gamma \left(\widetilde{\nu}_{\tau}
\rightarrow 
 \tau\tau\right) 
\approx\frac{m_{\widetilde{\nu}_{\tau}
}}{{16\pi}}
\left(Y_{\tau} Z^{H/A}_
{\widetilde{\nu}_{\tau} H_d}
- Y_{\nu_{\tau}} 
\frac{Y_{\tau}
}{\lambda}
\right)^2,
  \label{eq:3.220}
\end{equation}
where $Y_{\tau}\equiv Y_{e_{33}}$, 
and $Z^{H/A}$ is the matrix which diagonalizes the mass matrix for the
scalar/pseudoscalar Higgses ($H_d, H_u, \widetilde\nu_{R},
\widetilde\nu_{i}$)~\cite{Escudero:2008jg}. 
The latter is
determined by the neutrino Yukawas, which are the order parameters of
the $\rpv$. The contribution of $\lambda$ in the second term of
Eq.~(\ref{eq:3.220}) is due to the charged Higgsino mass that can be
approximated by the value of $\mu$. The partial decay width into $\tau
\ell$ can then be approximated for both sneutrino states by the second
term of Eq.~(\ref{eq:3.220}) with the substitution
$Y_{\nu_{\tau}}\rightarrow Y_{\nu_{\ell}}$: 
\begin{equation}
  \Gamma \left(\widetilde{\nu}_{\tau}
\rightarrow 
\tau \ell\right) 
\approx\frac{m_{\widetilde{\nu}
_{\tau}
}}{{16\pi}}
\left(Y_{\nu_{\ell}}
\frac{Y_{\tau} 
}{\lambda}
\right)^2.
  \label{eq:3.22}
\end{equation}

{On the other hand, 
the gauge interactions of 
$\widetilde{\nu}_{\tau}$ with neutrinos and Binos (Winos)
can produce also a large decay width into neutrinos,
via the gauge mixing between these gauginos and neutrinos.}
This partial decay width
can be approximated for scalar and pseudoscalar sneutrinos as
%
%
\begin{equation}
\sum_i \Gamma \left(\widetilde{\nu}_{\tau}
\rightarrow  
 \nu_\tau \nu_i\right)
\approx\frac{m_{\widetilde{\nu}_{\tau}}}{16\pi}
\frac{1}{2M^2} \sum_i{v_{i}^2},
  \label{eq:3.55}
\end{equation}
%
with $M$ a kind of average of Bino and Wino masses defined in 
Eq.~(\ref{eq:3.550}).
The relevant diagrams for  
$\widetilde{\nu}_{\tau}$
searches that include this decay mode are the same as in
Fig.~\ref{fig:production31}, but 
substituting one of the $\tau\,\ell/\tau$
vertices by a two-neutrino vertex.

Let us remark that other decay channels of 
the $\widetilde{\nu}_{\tau}$ can be present and have been taken into account in our numerical computation, but they turn out to be negligible
for the sneutrino masses that we are interested in this work.
{For example, decay to bottoms can occur through a term similar to the first one 
of Eq.~(\ref{eq:3.220}) with the substitution of $Y_{\tau}$ by $Y_b$. As we will comment below and was discussed in Ref.~\cite{Ghosh:2017yeh}, this term is very small. In particular, it is negligible with respect to the second one in Eq.~(\ref{eq:3.220}) which is present for decays into leptons.}

It is also worth noticing here that because the $\widetilde{\nu}_{\tau}$ in the $\mn$ has several relevant decay modes, the LEP lower bound on the sneutrino mass of
about 90 GeV~\cite{Abreu:1999qz,Abreu:2000pi,Achard:2001ek,Heister:2002jc,Abbiendi:2003rn,Abdallah:2003xc}
obtained under the assumption of BR one to leptons, via trilinear 
$\rpv$ couplings, is not directly applicable in this case. We have checked that no constraint on the $\widetilde{\nu}_{\tau}$ mass is obtained from these searches in the cases studied in this work.
We have obtained similar conclusions from LEP mono-photon search (gamma+MET)~\cite{Abdallah:2003np}, and LHC mono-photon and mono-jet (jet+MET) searches~\cite{Aaboud:2016uro,Aaboud:2016tnv}.
Concerning LEP searches for staus~\cite{Abreu:1999qz,Abreu:2000pi,Achard:2001ek,Heister:2002jc,Abbiendi:2003rn,Abdallah:2003xc}, in the $\mn$ the left stau does not decay directly but through an off-shell $W$ and a $\widetilde{\nu}_{\tau}$. Thus, searches for its direct decay are not relevant in this model. On the other hand, {the sneutrino mass can in principle be constrained using searches for final states as those produced in the $\mn$. However, we have checked that this is not the case. For example, for the final state $\tau\mu\mu\nu\tau\mu\mu\nu$ (see Fig.~\ref{fig:production31c}), taking into account the value of the production cross section at LEP for a pair of left staus, and the BRs of $W$ into $\mu\nu$ and $\widetilde{\nu}_{\tau}$ into $\tau\mu$, no effective constraint is obtained in our scenario. For the other possible topologies, with $W$ into quarks or $\widetilde{\nu}_{\tau}$ into $\nu\nu$, the results of the analyses turn out to be the same, as shown in Table~\ref{staustable}.
It is straightforward to see there that our scenario is unconstrained, even considering the most disfavored (and unrealistic) values for the branching ratios, BR($\widetilde{\nu}\rightarrow \ell\tau$)=1 or BR($\widetilde{\nu}\rightarrow 2\nu$)=1. In addition, we are not taking into account the possible effect of the different geometry of the decays when comparing with the final states considered in the searches, which are not originated in the same manner.}


\begin{table}[t]
 \caption{Possible topologies emerging from the production of a pair of tau left sneutrinos from staus. In the last two columns the production cross section of the most constrained signal, considering the worst-case scenario, is compared 
with the experimental upper limit. The values
BR($W\rightarrow \mu\nu$)=0.1063, BR($W\rightarrow q q'$)=0.6741, and BR($\tau \rightarrow \mu\nu$)=0.1739 are used for the computation, as well as 1/4 as the maximum value of BR($\tilde{\nu}\rightarrow \ell\tau$)$\times$ BR($\tilde{\nu}\rightarrow 2\nu$). In order to simplify the notation, BR($\tilde{\nu}\rightarrow \ell\tau$) means the sum of the three BRs to $e\tau$, $\mu\tau$ and $\tau\tau$, and the factors 1/3 are coming from considering the different channels.}
\label{staustable}
{\footnotesize
\begin{minipage}[b]{0.45\linewidth}\centering
  \begin{tabular}{*4c}
 \toprule
 Process& Topology & Signal cross section upper limit & Exclusion\\
 \midrule
  \midrule
 $\tilde{\tau}^{\pm}\tilde{\tau}^{\mp}\to 2 (W^{\pm}\to\ell^{\pm}\nu)$&4$\ell+2\tau+ \met$&$(0.13\ \mathrm{ pb})\times (0.1063)^2 $&
Fig.~14 of Ref.~\cite{Abbiendi:2003rn}\\  
 $+2(\tilde{\nu}\to\ell \tau)$&&$\times \mathrm{BR}(\tilde{\nu}\to\ell\tau)^2\times\left(\frac{1}{3}\right)^2 $& $1.5\times10^{-2}$ pb\\
 &&$\leq1.6\times10^{-4}\ \mathrm{ pb}$&\\\hline
 $\tilde{\tau}^{\pm}\tilde{\tau}^{\mp}\to 2 (W^{\pm}\to\ell^{\pm}\nu)$&3$\ell+\tau+ \met$&$(0.13\ \mathrm{ pb})\times (0.1063)^2 \times 2$&
Fig.~18 of Ref.~\cite{Abbiendi:2003rn}\\  
 $+(\tilde{\nu}\to\ell \tau)+(\tilde{\nu}\to2\nu)$&&$\times \mathrm{BR}(\tilde{\nu}\to\ell\tau)\times\frac{1}{3}\times(0.1739)$&$2\times10^{-2}$ pb\\
 &&$\times\mathrm{BR}(\tilde{\nu}\to2\nu)\leq4.3\times10^{-5} \ \mathrm{ pb}$&\\\hline
  $\tilde{\tau}^{\pm}\tilde{\tau}^{\mp}\to 2 (W^{\pm}\to\ell^{\pm}\nu)$&2$\ell+ \met$&$(0.13\ \mathrm{ pb})\times (0.1063)^2$&
Fig.~6 of Ref.~\cite{Abbiendi:2003rn}\\  
 $+2(\tilde{\nu}\to2\nu)$&&$ \times \mathrm{BR}(\tilde{\nu}\to2\nu)^2\leq1.5\times10^{-3} \ \mathrm{ pb}$&$6\times10^{-2}$ pb\\\hline
  $\tilde{\tau}^{\pm}\tilde{\tau}^{\mp}\to (W^{\pm}\to\ell^{\pm}\nu)$&3$\ell+2\tau+ jets+\met$&$(0.13\ \mathrm{ pb})\times2\times (0.1063)$&
Fig.~18 of Ref.~\cite{Abbiendi:2003rn}\\  
 $+(W^{\pm}\to q q')+2(\tilde{\nu}\to\ell \tau)$&&$\times(0.6741)\times \mathrm{BR}(\tilde{\nu}\to\ell\tau)^2 $&$2\times10^{-2}$ pb\\
 &&$\times\left(\frac{1}{3}\right)^2\leq2.1\times10^{-3}$ pb &\\\hline
   $\tilde{\tau}^{\pm}\tilde{\tau}^{\mp}\to (W^{\pm}\to\ell^{\pm}\nu)$&2$\ell+\tau+ jets+\met$&$(0.13\ \mathrm{ pb})\times2\times (0.1063)$&
Fig.~12 of Ref.~\cite{Abbiendi:2003rn}\\  
 $+(W^{\pm}\to q q')+(\tilde{\nu}\to\ell \tau)$&&$\times(0.6741)\times 2\times\mathrm{BR}(\tilde{\nu}\to\ell\tau)$&$5\times10^{-2}$ pb\\
 $+(\tilde{\nu}\to2\nu)$&&$\times\left(\frac{1}{3}\right)
\times\mathrm{BR}(\tilde{\nu}\to2\nu)$&
\\\
&&$\leq3.1\times10^{-3} \ \mathrm{ pb}$&\\\hline
   $\tilde{\tau}^{\pm}\tilde{\tau}^{\mp}\to (W^{\pm}\to\ell^{\pm}\nu)$&$\ell+ jets+\met$&$(0.13\ \mathrm{ pb})\times2\times (0.1063)$&
Fig.~6 of Ref.~\cite{Abbiendi:2003sc}\\  
 $+(W^{\pm}\to q q')+2(\tilde{\nu}\to2\nu)$&&$\times(0.6741)\times \mathrm{BR}(\tilde{\nu}\to2\nu)^2$&$5\times10^{-2}$ pb\\
 &&$\leq1.9\times10^{-2} \ \mathrm{ pb}$&\\\hline
  $\tilde{\tau}^{\pm}\tilde{\tau}^{\mp}\to 2 (W^{\pm}\to q q')$&2$\ell+2\tau+ jets$&$(0.13\ \mathrm{ pb})\times (0.6741)^2$&
Fig.~12 of Ref.~\cite{Abbiendi:2003rn}\\  
 $+2(\tilde{\nu}\to\ell \tau)$&&$ \times \mathrm{BR}(\tilde{\nu}\to\ell\tau)^2\times\left(\frac{1}{3}\right)^2$& $5\times10^{-2}$ pb\\
 &&$\leq6.6\times10^{-3} \ \mathrm{ pb}$&\\\hline
   $\tilde{\tau}^{\pm}\tilde{\tau}^{\mp}\to 2 (W^{\pm}\to q q')$&$\ell+\tau+ jets+\met$&$(0.13\ \mathrm{ pb})\times (0.6741)^2$&
Fig.~12 of Ref.~\cite{Abbiendi:2003rn}\\  
 $+(\tilde{\nu}\to\ell \tau)+(\tilde{\nu}\to2\nu)$&&$ \times2\times \mathrm{BR}(\tilde{\nu}\to\ell\tau)\times\left(\frac{1}{3}\right)$& $5\times10^{-2}$ pb\\
 &&$\times(0.1739)\times\mathrm{BR}(\tilde{\nu}\to2\nu)$ &\\
 &&$\leq1.7\times10^{-3}$ pb &\\\hline
   $\tilde{\tau}^{\pm}\tilde{\tau}^{\mp}\to 2 (W^{\pm}\to q q')$&$jets+ \met$&$(0.13\ \mathrm{ pb})\times (0.6741)^2$&
Fig.~20 of Ref.~\cite{Abbiendi:2003rn}\\  
 $+2(\tilde{\nu}\to2\nu)$&&$ \times \mathrm{BR}(\tilde{\nu}\to2\nu)\leq6\times10^{-2} \ \mathrm{ pb}$&$5\times10^{-1}$ pb\\
      \bottomrule
     \end{tabular}
 \end{minipage}
 }
\end{table}

To analyze now the BRs into leptons, we have to focus on the decay channels where the
$\tau$'s in the final state decay leptonically. {Let us study e.g. the BR to $\mu\mu$, since the $ee$ channel can be discussed in a similar way, and the BR to $e\mu$ fulfills
$\text{BR}(\widetilde{\nu}_{\tau} \rightarrow e \mu)\approx
\text{BR}(\widetilde{\nu}_{\tau} \rightarrow \mu \mu) + 
\text{BR}(\widetilde{\nu}_{\tau} \rightarrow e e)$
given that the BRs of the 
$\tau$ decays into electrons or muons (plus neutrinos) are similar $\approx$ 0.17.
To quantify roughly the value 
$\text{BR}(\widetilde{\nu}_{\tau} \rightarrow \mu\mu)$,
we can use the following formula:
%
\begin{equation}
\text{BR} \left(\widetilde{\nu}_{\tau}
\rightarrow  \mu \mu\right)
 \approx 0.068\times
\biggl(1+ \frac{r}{3} \biggr)^{-1},
  \label{eq:3.2211}
\end{equation}
with
\begin{equation}
r\approx\left(\frac{\lambda}{Y_{\tau}}\right)^2 \frac{2m_{\nu}}{Y_{\nu}^2 M},
\label{neuvsl}
\end{equation}
%
where we have neglected the first term in Eq.~(\ref{eq:3.220}), which is a sensible approximation for small sneutrino masses 
around $50$ GeV provided that $\lambda$ is not large,
and we have used the neutrino mass formula discussed below Eq.~(\ref{eq:3.550}), 
$m_{\nu} \approx {\sum_i {v_{i}^2}}/{4M}$, implying that the decay width in 
Eq.~(\ref{eq:3.55}) can be written as
$(m_{\widetilde{\nu}_{\tau}}/{16\pi}) 2m_{\nu}/M$.
In addition, we have
set all neutrino Yukawas with a common value $Y_{\nu}$ in
order to simplify the analysis. In what follows we will continue with this strategy, which does not essentially modify the
results. Now, in the above equations and 
for typical values of the parameters such as e.g. $Y_{\nu} = 5\times 10^{-7}$, $m_{\nu} = 0.05$ eV and $M = 1$ TeV, one obtains $r = 0.4$ 
and therefore a 
$\text{BR} (\widetilde{\nu}_{\tau}
\rightarrow  \mu \mu ) \lsim 0.06$
for $\lambda\gsim Y_{\tau}$.}

{In this approximation,
we can also write the proper decay distance as
\begin{equation}
c\tau\approx 0.22 \times  \biggl(\frac{Y_{\nu}}{5\times 10^{-7}}\biggr)^{-2} \left(\frac{\lambda}{Y_{\tau}}\right)^2
\biggl(1+ \frac{r}{3} \biggr)^{-1}
\biggl(\frac{m_{\widetilde{\nu}_{\tau}}}{60~\text{GeV}}\biggr)^{-1} ~ \text{mm},
 \label{eq:3.22112}
\end{equation}
obtaining
$c\tau\gsim 0.2$ mm for $\lambda\gsim Y_{\tau}$.
Thus the latter is a necessary condition on $\lambda$ in order to obtain suitable displaced vertices. In fact, as we will see in the next sections, 
we will need decay lengths larger than about a millimeter in order to be constrained by the current experimental results. 
}


\begin{figure}[t]
   \centering
       \subcaptionbox{\label{fig:BRa}
       {Branching ratio
}
}
{\epsfig{file=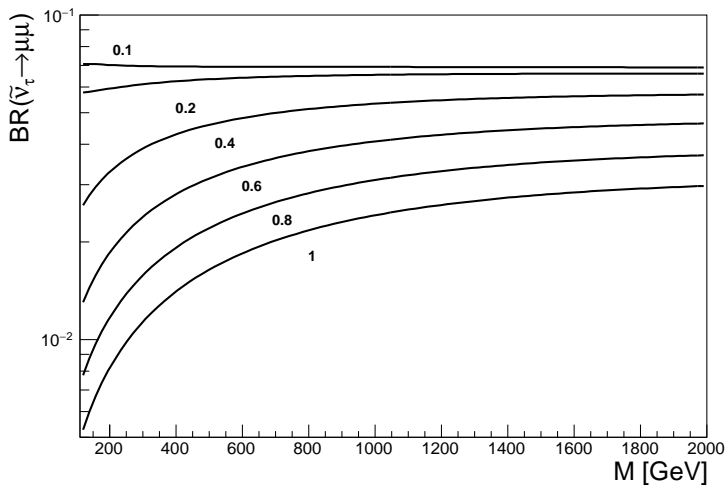,height=7cm}}
 \subcaptionbox{\label{fig:BRb}{Decay length 
}}
  {\epsfig{file=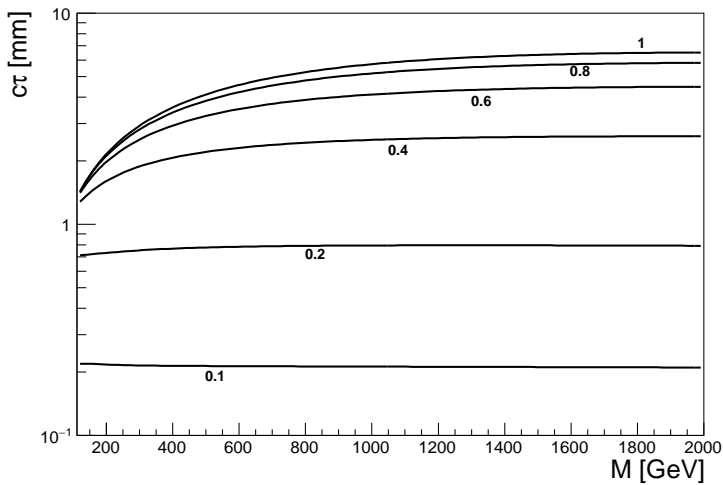,height=7cm}}
\caption{
(a) {BR versus $M$ for the decay of a 
$\widetilde{\nu}_{\tau}$
with $m_{\widetilde{\nu}_{\tau}} = 60$ GeV into $\mu\mu$};
(b) Proper decay distance $c\tau$ of the $\widetilde{\nu}_{\tau}$ versus $M$. {In both plots (a) and (b),  the neutrino Yukawas are set to $Y_\nu = 5\times 10^{-7}$, and several values of the coupling $\lambda$ are used such as $\lambda=$ 0.1, 0.2, 0.4, 0.6, 0.8, 1.}  
}
     \label{fig:BR}
\end{figure}

To compute numerically
the mass spectrum and decay modes, we used a suitable modified version of the 
{\tt SARAH} code~\cite{Staub:2008uz} as well as the {\tt SPheno} 3.3.6 code~\cite{Porod:2003um,Porod:2011nf}. 
As an example, we show
in Fig.~\ref{fig:BRa} the BR for the decay of a scalar 
sneutrino $\widetilde{\nu}^\mathcal{R}_{\tau}$ with a mass
of 60 GeV into $\mu\mu$, {for $Y_\nu = 5\times 10^{-7}$}
(a similar figure is obtained in the case of the pseudoscalar 
$\widetilde{\nu}^\mathcal{I}_{\tau}$).
This is plotted 
as a function of $M$ 
for several values of the coupling $\lambda$. Values of $M$ smaller than
111.3~GeV are not considered since the $\widetilde{\nu}_{\tau}$
would no longer be the LSP in favor of the gauginos.
{Other parameters, whose effect on the sneutrino decay properties is less
significant, as can be understood from previous formulas, are set to be $\kappa=0.3$, $\tan\beta=10$, and $\frac{v_R}{\sqrt 2}= 1350$ GeV,
throughout this work. 
Concerning the quantity $\sum_iv_{i}^2$ in Eq.~(\ref{eq:3.55}), this is determined 
using Eq.~(\ref{neumass}) with the heavier neutrino mass fixed by the experimental constraints in 
the 
range $m_\nu \sim$ [0.05, 0.23] eV, i.e. below the upper bound on the sum of neutrino masses $\sim 0.23$ eV \cite{Ade:2015xua}, and above the square root of the mass-squared difference 
$\Delta m_{\rm atm}^2\sim 2.42\times 10^{-3}\mathrm{eV}^2$ \cite{An:2015rpe}.
In Fig.~\ref{fig:BR}a we chose as an example $m_\nu = 0.05$ eV. 
Finally, given the 
sneutrino mass formula in Eq.~(\ref{evenLLL2}),  
there is enough freedom to tune 
$A_{{\nu}_i}$ in order to get the $\widetilde{\nu}_{\tau}$ as the LSP
with a mass of 60 GeV as in the case of Fig.~\ref{fig:BR}a. 
We can see in the figure that 
small values of $\lambda$ and large values of
$M$ favor larger BRs to dileptons.
These results can be easily deduced from Eqs.~\eqref{eq:3.2211} and \eqref{neuvsl}.}

In Fig.~\ref{fig:BRb}, 
we show the proper decay distance $c\tau$ of the 
$\widetilde{\nu}^\mathcal{R}_{\tau}$ for the same values of the parameters as 
in Fig.~\ref{fig:BRa}.
Large values of $\lambda$ and
$M$ favor larger decay lengths, as can be understood 
from Eqs.~\eqref{eq:3.22112} and \eqref{neuvsl}. 
As mentioned before, we need 
decay lengths larger than about a millimeter, and therefore for these values of the parameters 
the coupling is constrained to be $\lambda \gsim 0.2$. 
For example, for $\lambda = 1$ we can see
that the upper bound on the decay length is $c\tau = 7$ mm. 
However, large values of $\lambda$ also favor smaller BRs into leptons, and therefore less stringent constraints on the parameter space. The interplay between these effects will be analyzed in the next sections.
In addition, the values of $Y_{\nu}$ and $m_{\widetilde{\nu}_{\tau}}$ also play an important role in the analysis.
We can see in Eq.~(\ref{eq:3.22112}) that smaller values favor larger decay lengths.
For example, for $m_{\widetilde{\nu}_{\tau}} = 60$ GeV as in Fig.~\ref{fig:BR}
but $Y_\nu = 10^{-7}$, we have checked that the upper bound on the decay length is 
$c\tau = 20$ mm for the case of $\lambda=1$. For $Y_\nu = 5\times 10^{-7}$
as in Fig.~\ref{fig:BR} but $m_{\widetilde{\nu}_{\tau}} = 80$ GeV, the upper bound 
for $\lambda=1$ turns out to be smaller as expected, $c\tau = 4$ mm.
{Let us finally remark that $Y_{\nu}$
cannot be made arbitrarily small. 
Because of the relation $v_i\sim Y_{\nu} v_u$ discussed in the Introduction,
this would imply that $v_i$ has to be also small coming into 
conflict with 
Eq.~(\ref{neumass}) and the constraint discussed above, $m_\nu \sim$ [0.05, 0.23] eV. Thus a reasonable range for $Y_{\nu}$ is between $10^{-6}$ and $10^{-7}$}.

\section{Long-lived particle searches at the LHC}
\label{reinterpretation}
 
As discussed in the previous section, a tau left sneutrino $\widetilde{\nu}_{\tau}$ can decay
into a pair of leptons with a proper lifetime of $\gtrsim
100~\mu\mathrm{m}$ (see Fig.~\ref{fig:BRb}), long enough to have a
visible separation from the production point. This signal can be
compared with the searches for long-lived particles at the LHC.  

There are various long-lived particle searches at the LHC, and
each of them aims at a signature specific to a particular kind of
particles. We thus first need to discuss which search strategy is most
sensitive to $\widetilde{\nu}_{\tau}$. Since they are
electrically neutral, we are unable to use disappearing track searches
or metastable charged particle searches to probe them. On the other
hand, as we have seen above, the decay products of the sneutrinos include
charged particles, and therefore we may detect the longevity of the
sneutrinos by reconstructing their decay vertices, using the
charged tracks associated with the daughter particles. This type of
search strategies is dubbed as the displaced-vertex searches. 

Both the ATLAS and CMS experiments have been searching for displaced
vertices. The ATLAS 8-TeV analysis \cite{Aad:2015rba} searches for
events containing at least one long-lived particle decaying at a
significant distance from the production point ($\gtrsim 1$~mm), looking
for decays into two leptons or into five or more charged particles. The
latter search channel focuses on processes that produce a higher amount
of charged particles and/or missing energy, compared with the decay of
the $\widetilde{\nu}_{\tau}$. We will however see below that the
dilepton search channel in the ATLAS 8-TeV analysis is general enough as
to be sensitive to the decay of the $\widetilde{\nu}_{\tau}$.  
On the other hand, the current 13-TeV
displaced-vertex search performed by the ATLAS collaboration
\cite{Aaboud:2017iio} is optimized for long-lived gluinos, so it is not
possible to use it for our purpose. As for the CMS, the 8-TeV analysis
presented in Ref.~\cite{CMS:2014hka} gives a limit comparable to the
ATLAS 8~TeV bound if the decay distance of sneutrinos is $\gtrsim
3$~cm---although this CMS search is in principle sensitive to shorter
decay distances as the selection cut requires $|d_0| < 12\sigma_d \sim
180~\mu\mathrm{m}$ ($d_0$ is the transverse impact parameter of tracks and
$\sigma_d$ is its uncertainty), their limits are terminated at a much
larger value of $c\tau$ especially for soft displaced vertices (see,
{\it e.g.}, Fig.~6 in Ref.~\cite{CMS:2014hka}). Given that the ATLAS
8-TeV analysis provides bounds for smaller $c\tau$ compared with those
from the CMS study, to make the discussions concrete, we focus on
displaced-vertex searches with the ATLAS detector in what follows.

The ATLAS displaced-vertex search in Ref.~\cite{Aad:2015rba} is based on
the 8-TeV data with an integrated luminosity of 20.3~fb$^{-1}$. Among
the various search channels studied in the analysis, the dilepton
displaced-vertex selection channel, where each displaced vertex is 
formed from at least two oppositely-charged leptons, may be used for the
long-lived $\widetilde{\nu}_{\tau}$
search. 
As we mentioned above, we focus on the
decay processes of 
$\widetilde{\nu}_{\tau}$ 
in which $\tau$ leptons in the final
state decay into leptons, in order to utilize this selection channel.

In the dilepton displaced-vertex search, each event must satisfy
the muon or electron trigger requirement.\footnote{In the ATLAS search,
the missing-energy and jets triggers are also used. These triggers are
ineffective in our setup. } For the muon trigger, a muon
candidate is identified only in the muon spectrometer, without utilizing
the tracking information, and required to
have a transverse momentum of $p_{\rm T} > 50$~GeV and the
pseudorapidity of $|\eta| < 1.07$. For the electron trigger,
only a high-energy deposit in the electromagnetic calorimeter is
required, again without tracker requirements. This has a less effective
background rejection performance compared with the muon trigger, and
thus a relatively strong criterion is imposed on the transverse momentum
of electrons: either a single electron with $p_{\rm T} > 120$~GeV or two
electrons with $p_{\rm 
T} > 40$~GeV. The events which have passed these triggers are then
required to be subject to the object reconstruction and
filtering criteria. Finally, with the help of the retracking procedure,
a dilepton displaced vertex is reconstructed from two oppositely-charged
lepton tracks: $\mu^+ \mu^-$, $e^+ e^-$, or $e^{\pm} \mu^{\mp}$. Here,
the lepton tracks are required to satisfy $p_{\rm T} > 10$~GeV, $0.02
\leq |\eta| < 2.5$, and $d_0 > 2$~mm (2.5~mm) for muons (electrons). In addition, the
invariant mass of the tracks, $m_{\rm DV}$, should be larger than
10~GeV. The position of
the reconstructed displaced vertices must satisfy $r_{\rm DV} < 300$~mm,
$|z_{\rm DV}| < 300$~mm, and $\sqrt{(x_{\rm DV}-x_{\rm PV})^2 + (y_{\rm
DV} -y_{\rm PV})^2} \geq 4$~mm, where the subscripts DV and PV indicate
that the corresponding coordinates are those of the displaced vertex and
the primary vertex, respectively. The effect of the first two conditions
is almost negligible in our analysis since the decay distance of
sneutrinos is $\lesssim 10$~mm as shown in Fig.~\ref{fig:BRb}, while the
third condition does affect our analysis, as we shall see below. 

With these requirements, the ATLAS collaboration searched for dilepton
displaced vertices and found no event, while the numbers of background
events are expected to be $1.0\pm 0.2^{+0.3}_{-0.6} \times 10^{-3}$,
$2.4\pm 0.9^{+0.8}_{-1.5} \times 10^{-3}$, and $2.0\pm 0.5^{+0.3}_{-1.4}
\times 10^{-3}$ for the $e^+ e^-$, $e^\pm \mu^\mp$, and $\mu^+ \mu^-$
channels, respectively. The dominant source of the background is
accidental crossings of independent lepton tracks. 
As we see, this search is basically background
free. With this result, strong limits were imposed on long-lived
particles which decay into leptons.

We however cannot directly apply the limits provided by the ATLAS
collaboration (Fig.~13 of Ref.~\cite{Aad:2015rba}) to the left sneutrino
case, since the ATLAS analysis simulates the decay of a heavy gluino
into a light and a heavy neutralino. The former case represents a highly
boosted light particle decaying into a pair of muons, while the latter
represents a heavy non-boosted particle decaying in the same way. Yet,
the sneutrino features a light non-boosted particle. This analysis can be
extended, nevertheless, combining information from both
situations by considering the fact that the difference in the strength
of the upper limits basically comes from the efficiency in passing the
event selection requirements (the decay products from a
sufficiently heavy neutralino are so energetic that almost all the
events pass the selection criteria),\footnote{In the ATLAS search
\cite{Aad:2015rba}, the light neutralino events can pass the trigger
requirement since the missing-energy and jets triggers are also used,
but the vertex-level efficiency deteriorates for such
events. With keeping this in mind, we here assume that all of the neutralino
events satisfy the trigger requirements---the dilepton trigger for the heavy
case and the missing-energy/jets trigger for the light case---and the
difference in the sensitivities originates from that in the vertex-level
efficiencies. This assumption is fairly reasonable since the leptons
(jets) in the final state are very active in the heavy (light)
neutralino case. 
} while the position of the
minimum of the limits is determined by the boost factors of the
neutralinos. Thus, what we can do is shifting upwards the limit
corresponding to the non-boosted neutralino to make its minimum to
coincide with the one of the line corresponding to the light boosted
neutralino.\footnote{We however note that there is also an efficiency loss for
a boosted system as the final state muons tend to be collinear with each
other, whose effect is not taken into account in this
prescription. The neglect of this effect thus results in a rather
conservative limit.} The resultant limit for the dimuon channel is
displayed as a function of the decay distance $c\tau$ in the
purple-shaded solid line in Fig.~\ref{Fig13}. We also show the limits
corresponding to the light boosted and heavy non-boosted cases in the green-hatched
dotted and yellow-hatched dash-dotted lines, respectively, which are taken from
Fig.~13 (c) of Ref.~\cite{Aad:2015rba}.  
{We analyzed the limits
for the $ee$ and $e\mu$ channels in a similar manner using Figs.~13 (a) and
(b) in Ref.~\cite{Aad:2015rba}, respectively, obtaining very similar plots}. 
As seen from Fig.~\ref{Fig13}, the ATLAS
displaced-vertex search is sensitive to a decay distance larger than about a few mm. 
This stems from the requirements that the impact parameter $d_0$ of the
muon tracks be larger than 2~mm and the transverse distance between the
displaced vertices and the primary vertices be larger than 4~mm, as we
mentioned above. We will discuss a possibility of relaxing these
requirements later.

\begin{figure}[t!]
\centering
\includegraphics[scale=0.50]{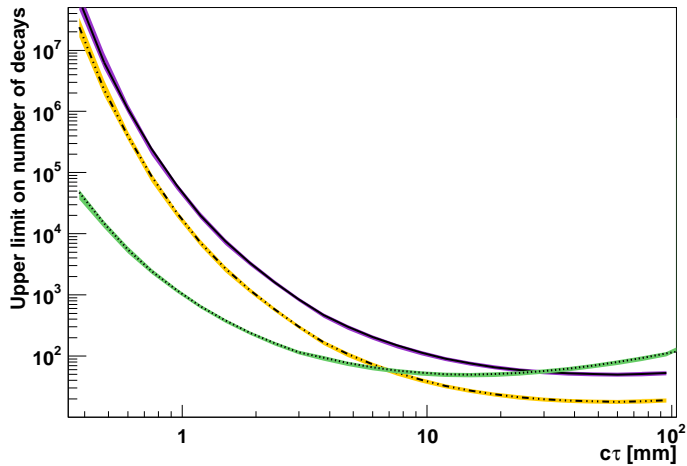} 
\caption{{Upper limit at 95$\%$ confident level on the number of tau left sneutrinos
decaying into dimuons at $\sqrt s = 8$ TeV for an integrated luminosity of 20.3
 fb$^{-1}$, shown in the purple-shaded solid line. We also show the
 limits given in Fig.~13 (c) of Ref.~\cite{Aad:2015rba}, which we use to
 obtain our limit: green-hatched dotted (yellow-hatched dash-dotted) line
corresponding to a light boosted neutralino of 50 GeV 
(heavy non-boosted neutralino of 1000 GeV) from the decay of a heavy gluino of 1300 GeV.
}  
}
\label{Fig13}
\end{figure} 

Another obstacle for the left sneutrino case is the trigger
requirement. Since left sneutrinos we consider in this paper have a
mass of $\lesssim 100$~GeV and are less boosted, their decay products
have relatively small momenta. On the other hand, the ATLAS 8-TeV
analysis requires rather high thresholds for lepton momenta,
especially for electrons,  since it aims at generic long-lived
particles such as metastable neutrinos produced by the decays of colored
particles. In particular, the events must satisfy the following requirements~\cite{Aad:2015rba}:
\begin{itemize}
 \item[$\bullet$] One muon with $p_{\rm T}>50$~GeV and
$|\eta|<1.07$, one electron
with $p_{\rm T}>120$~GeV or two electrons with $p_{\rm T}>40$~GeV.
 \item[$\bullet$] One pair $e^+e^-$, $\mu^+\mu^-$ or $e^\pm\mu^\mp$ 
with $p_{\rm T} >10$~GeV and $0.02<|\eta|<2.5$ for each one.
\end{itemize}
To estimate the sensitivity of this search strategy, samples of 
simulated Monte Carlo (MC) events are used to study the efficiency of
the triggering and off-line selection processes for signal events. In
each event two sneutrinos are created in the $pp$ collision as described
in Section~\ref{phenomenology}. 
All simulated samples are generated using {\tt MadGraph5\_aMC@NLO} 2.6.0
\cite{Alwall:2014hca} and {\tt PYTHIA} 8.230 \cite{Sjostrand:2007gs}. 
{In particular, ten thousand events are generated for each case 
with {\tt MadGraph5\_aMC@NLO} 2.6.0 at leading order (LO) of perturbative QCD simulating 
the production of the described process. 
We include the next-to-leading order (NLO)~\cite{Baer:1997nh} and next-to-leading logarithmic accuracy (NLL)~\cite{Fuks:2013lya} results using a $K$-factor of about 1.2.}
We also use \texttt{DELPHES v3} \cite{deFavereau:2013fsa} for the
detector simulation. 
The effect of these selections for different masses and for the three production processes 
at $\sqrt s = 8$ TeV, 
is shown in Tables~\ref{new} and~\ref{new2}
for the
$\mu\mu$ and $\mu e$ channels, respectively.
The trigger requirement for electrons is too restrictive and makes the selection efficiency for the dielectron channel to be a few percent level, and thus ineffective for light sneutrino searches. 
In these tables, $\epsilon_{\rm sel}$ is the
selection efficiency for each case.
Using these results and the BRs discussed in Sec.~\ref{phenomenology},
we can compute the prediction for the number of decays
$\widetilde{\nu}_{\tau}\to\mu\mu /\mu e$ passing the trigger and event
selection requirements. Notice however that $\epsilon_{\rm sel}$ is not large.


\begin{table}[t]
 \caption{Cutflow of the selection requirements for
 $\widetilde{\nu}_{\tau}\to\mu\mu$ for the 8-TeV analysis,
where $\epsilon_{\rm sel}$ is the
selection efficiency. {For each case, 10,000 MC events are generated.}}
\label{new}
{\footnotesize
\begin{minipage}[b]{0.45\linewidth}\centering
  \begin{tabular}{c|*4c}
 \toprule
&\multicolumn{4}{c}{ Through $Z$ boson }\\
 \midrule
  $m_{\widetilde{\nu}_{\tau}}$ (GeV)&50&60&80&100\\  \hline
  $p_{\rm T_{\text{trigger}}}$&2,280&4,241&6,828&8,063\\  
  dilepton sel. &1,069&2,096&3,517&4,333\\  
   \bottomrule
  $\epsilon_{\text{sel}}$&0.1069&0.2096&0.3517&0.4333\\  \hline\hline
 \end{tabular}
 \end{minipage}
\hspace{-0.5cm}
\begin{minipage}[b]{0.45\linewidth}
\centering
  \begin{tabular}{|*4c}
 \toprule
 \multicolumn{4}{|c}{Through $W$ boson}\\
 \midrule
  50&60&80&100\\  \hline
  2,335&3,703&3,840&7,515\\  
  1,135&1,881&3,740&3,968\\
    \bottomrule
  0.1135&0.1881&0.374&0.3968\\  \hline\hline
 \end{tabular}
\end{minipage}
}
{\begin{center}
\footnotesize
\begin{minipage}[b]{0.45\linewidth}\centering
  \begin{tabular}{c|*4c}
 \toprule
&\multicolumn{4}{c}{ Through $\gamma ,Z$ bosons}\\
 \midrule
&50&60&80&100\\  \hline
&1,616&2,233&3,396&4,374\\  
&1,550&2,150&3,273&4,208\\  
   \bottomrule
&0.1550&0.2150&0.3273&0.4208\\  \hline\hline
 \end{tabular}
 \end{minipage}
 \end{center}
\hspace{1.4cm}
}
\end{table}
\begin{table}[t]
 \caption{
The same as in Table~\ref{new}
 but for $\widetilde{\nu}_{\tau}\to\mu e$.
}
\label{new2}
{\footnotesize
\begin{minipage}[b]{0.45\linewidth}\centering
  \begin{tabular}{c|*4c}
 \toprule
&\multicolumn{4}{c}{ Through $Z$ boson }\\
 \midrule
  $m_{\widetilde{\nu}_{\tau}}$ (GeV)&50&60&80&100\\  \hline
  $p_{\rm T_{\text{trigger}}}$&659&1,347&2,526&3,612\\  
  dilepton sel.&618&1,253&2,315&3,278\\  
   \bottomrule
  $\epsilon_{\text{sel}}$&0.0618&0.1253&0.2315&0.3278\\  \hline\hline
 \end{tabular}
 \end{minipage}
\hspace{-0.45cm}
\begin{minipage}[b]{0.45\linewidth}
\centering
  \begin{tabular}{|*4c}
 \toprule
 \multicolumn{4}{|c}{Through $W$ boson}\\
 \midrule
  50&60&80&100\\  \hline
  678&1,138&2,134&3,067\\  
  628&1,061&1,972&2,803\\
    \bottomrule
  0.0628&0.1061&0.1972&0.2803\\  \hline\hline
 \end{tabular}
\end{minipage}
}
{\begin{center}
\footnotesize
\begin{minipage}[b]{0.45\linewidth}\centering
  \begin{tabular}{c|*4c}
 \toprule
&\multicolumn{4}{c}{ Through $\gamma ,Z$ bosons}\\
 \midrule
&50&60&80&100\\  \hline
&878&1,249&2,329&3,287\\  
&827&1,160&2,109&2,978\\  
   \bottomrule
&0.0827&0.116&0.2109&0.2978\\  \hline\hline
 \end{tabular}
 \end{minipage}
 \end{center}
\hspace{1.4cm}
}
\end{table}

The ATLAS 8-TeV search strategy discussed above is, of course, not optimized for the present setup.
Nevertheless, in principle it is possible to optimize the trigger
requirements for left sneutrino searches by relaxing the thresholds,
as there are a variety of different lepton triggers with lower momentum
thresholds used in the ATLAS experiment. For instance, for the muon
trigger, the ATLAS 8-TeV analysis uses only the muon spectrometer and
requires $p_{\rm T} > 50$~GeV, as discussed above. On the other hand, the {\tt mu24i}
trigger \cite{Aad:2014sca}, which is an isolated single muon trigger at
the event-filter, also uses the information from the inner detector and
requires the transverse momentum threshold of $p_{\rm T} >
24$~GeV.\footnote{This 
trigger should also satisfy a loose isolation selection, the sum of the
$p_{\rm T}$ of tracks in a cone of $\Delta R<0.2$ centered around the
muon candidate after eliminating the muon transverse momentum
$(p_{\rm T})_\mu$ should be smaller than $0.12 \times (p_{\rm T})_\mu$;
this requirement is so loose that almost all isolated muons from the
$Z$-boson decays pass the criterion. Since the muons coming from the
sneutrino decays are also expected to be isolated, we can expect that
this requirement scarcely affects the sneutrino event selection. For
this reason, we do not take account this effect in the following
analysis.} 
With the help of the inner detector information, this {\tt mu24i}
trigger has a good performance in a wider range of the pseudorapidity of
tracks, and thus we can also relax the requirement on $\eta$; from
$|\eta| < 1.07$ to $|\eta| < 2.5$ \cite{Aad:2014sca}. To exploit this
trigger instead of that used in Ref.~\cite{Aad:2015rba} can
significantly enhance the sensitivity to light sneutrinos, since the
typical momentum of muons from the sneutrino decays is a few tens of
GeV. A side effect of the reduction of the momentum threshold is, of
course, an increase of the number of background events. According to
Ref.~\cite{Aad:2014sca}, the enhancement in the number of events due to
the relaxation of the trigger requirement is expected to be $\sim 10$. 
Since the main background in the displaced-vertex search is accidental
crossings of uncorrelated lepton tracks \cite{Aad:2015rba}, we can
estimate the increase in the number of background events by scaling this
according to the number of events passing the trigger. Given that the number of
background muon vertices in the ATLAS 8-TeV search is as low as $\sim 2 \times
10^{-3}$ \cite{Aad:2015rba}, we can safely conclude that the number of
background events can still be regarded as zero even if we relax the
trigger requirement for muons. Another restriction we need to take into
account is the requirement on the impact parameter $d_0$ of muon
tracks adopted by {\tt mu24i}; $|d_0| < 1$~cm is required
\cite{Aad:2012xs, Aad:2008zzm} for the {\tt mu24i} trigger, which indicates that the
efficiency should be reduced for sneutrinos with $c\tau \gtrsim
1$~cm. Nevertheless, this again causes a negligible effect on left
sneutrino searches in the present setup since the sneutrinos have a
proper decay distance smaller than 1~cm,\footnote{This is the reason why
we show only a small $c\tau$ region in Fig.~\ref{Fig13}. } as shown in
Fig.~\ref{fig:BRb}. We therefore conclude that the use of the {\tt
mu24i} trigger instead of the present one in Ref.~\cite{Aad:2015rba} is
very powerful and promising for the left sneutrino searches. 

\begin{table}[t]
 \caption{Cutflow of the selection requirements for
 $\widetilde{\nu}_{\tau}\to\mu\mu$ for the optimized 8-TeV analysis,
where $\epsilon_{\rm sel}$ is the
selection efficiency. {For each case, 10,000 MC events are generated.}}
\label{table:cutflow1}
{\footnotesize
\begin{minipage}[b]{0.45\linewidth}\centering
  \begin{tabular}{c|*4c}
 \toprule
&\multicolumn{4}{c}{ Through $Z$ boson }\\
 \midrule
  $m_{\widetilde{\nu}_{\tau}}$ (GeV)&50&60&80&100\\  \hline
  $p_{\rm T_{\text{trigger}}}$&6,687&7,775&8,682&9,144\\  
  dilepton sel. &6,637&7,707&8,589&9,027\\  
   \bottomrule
  $\epsilon_{\text{sel}}$&0.6637&0.7707&0.8589&0.9027\\  \hline\hline
 \end{tabular}
 \end{minipage}
\hspace{-0.45cm}
\begin{minipage}[b]{0.45\linewidth}
\centering
  \begin{tabular}{|*4c}
 \toprule
 \multicolumn{4}{|c}{Through $W$ boson}\\
 \midrule
  50&60&80&100\\  \hline
  6,520&7,343&7,359&8,846\\  
  6,409&7,198&7,229&8,594\\
    \bottomrule
  0.6409&0.7198&0.7229&0.8594\\  \hline\hline
 \end{tabular}
\end{minipage}
{\begin{center}
\footnotesize
\begin{minipage}[b]{0.45\linewidth}\centering
  \begin{tabular}{c|*4c}
 \toprule
&\multicolumn{4}{c}{ Through $\gamma ,Z$ bosons}\\
 \midrule
&50&60&80&100\\  \hline
&6,961&7,471&8,387&8,916\\  
&6,701&7,288&8,151&8,647\\  
   \bottomrule
&0.6701&0.7288&0.8151&0.8647\\  \hline\hline
 \end{tabular}
 \end{minipage}
 \end{center}
}
}
\end{table}
\begin{table}[t]
 \caption{The same as in Table~\ref{table:cutflow1}
 but for $\widetilde{\nu}_{\tau}\to\mu e$.
}
\label{table:cutflow2}
{\footnotesize
\begin{minipage}[b]{0.45\linewidth}\centering
  \begin{tabular}{c|*4c}
 \toprule
&\multicolumn{4}{c}{ Through $Z$ boson }\\
 \midrule
  $m_{\widetilde{\nu}_{\tau}}$ (GeV)&50&60&80&100\\  \hline
  $p_{\rm T_{\text{trigger}}}$&4,014&5,255&6,508&7,420\\  
  dilepton sel.&3,943&5,094&6,267&7,067\\  
   \bottomrule
  $\epsilon_{\text{sel}}$&0.3943&0.5094&0.6267&0.7067\\  \hline\hline
 \end{tabular}
 \end{minipage}
\hspace{-0.45cm}
\begin{minipage}[b]{0.45\linewidth}
\centering
  \begin{tabular}{|*4c}
 \toprule
 \multicolumn{4}{|c}{Through $W$ boson}\\
 \midrule
  50&60&80&100\\  \hline
  3,853&4,754&6,054&6,974\\  
  3,763&4,624&5,830&6,637\\  
   \bottomrule
  0.3763&0.4624&0.5830&0.6637\\  \hline\hline
 \end{tabular}
\end{minipage}
\begin{center}
\footnotesize
\begin{minipage}[b]{0.45\linewidth}\centering
  \begin{tabular}{c|*4c}
 \toprule
&\multicolumn{4}{c}{ Through $\gamma ,Z$ bosons}\\
 \midrule
&50&60&80&100\\  \hline
&6,118&6,748&7,116&8,293\\  
&4,917&5,515&6,360&6,961\\  
   \bottomrule
&0.4917&0.5515&0.6360&0.6916\\  \hline\hline
 \end{tabular}
 \end{minipage}
 \end{center}
}
\end{table}

We may also use a lower $p_{\rm T}$ threshold for the electron
trigger. However, we are unable to estimate the increase in the number
of background events in this case from, say,
Ref.~\cite{ATLAS:electrontrigger}, since the plot does not show the
corresponding trigger rate for $p_{\rm T} > 120$~GeV. Considering this,
in the following 8-TeV analysis, we only use the muon trigger with
$p_{\rm T}>24$~GeV and consider the $\mu^+\mu^-$ and $\mu^\pm
e^\mp$ channels to be conservative. We however note that we can
certainly optimize the electron trigger as well, which indeed improves the
sensitivity considerably and thus is worth a further dedicated study.

\begin{table}[t]
 \caption{Cutflow of the selection requirements for
 $\widetilde{\nu}_{\tau}\to\mu\mu$, for the optimized 13-TeV analysis,
where $\epsilon_{\rm sel}$ is the
selection efficiency. {For each case, 10,000 MC events are generated.}}
\label{table:cutflow3}
{\footnotesize
\begin{minipage}[b]{0.45\linewidth}\centering
  \begin{tabular}{c|*4c}
 \toprule
&\multicolumn{4}{c}{ Through $Z$ boson}\\
 \midrule
  $m_{\widetilde{\nu}_{\tau}}$ (GeV)&50&60&80&100\\  \hline
  $p_{\rm T_{\text{trigger}}}$
&5,797&7,011&8,077&8,581\\  
  dilepton sel.
&5,739&6,941&7,995&8,469\\  
   \bottomrule
  $\epsilon_{\text{sel}}$&0.5739&0.6941&0.7995&0.8469\\  \hline\hline
 \end{tabular}
 \end{minipage}
\hspace{-0.45cm}
\begin{minipage}[b]{0.45\linewidth}
\centering
  \begin{tabular}{|*4c}
 \toprule
 \multicolumn{4}{|c}{Through $W$ boson}\\
 \midrule
  50&60&80&100\\  \hline
  5,643&6,539&7,587&8,281\\  
  5,587&6,411&7,384&8,020\\  
     \bottomrule
  0.5587&0.6411&0.7384&0.8020\\  \hline\hline
 \end{tabular}
\end{minipage}
\begin{center}
\footnotesize
\begin{minipage}[b]{0.45\linewidth}\centering
  \begin{tabular}{c|
*4c}
 \toprule
&\multicolumn{4}{c}
{Through $\gamma ,Z$ bosons}\\
 \midrule
&50&60&80&100\\  \hline
&5,885&6,634&7,716&8,316\\  
&5,705&6,459&7,485&8,000\\  
   \bottomrule
&0.5705&0.6459&0.7485&0.8\\  \hline\hline
 \end{tabular}
 \end{minipage}
 \end{center}

}
\end{table}

\begin{table}[t]
 \caption{The same as in Table~\ref{table:cutflow3} but for
$\widetilde{\nu}_{\tau}\to\mu e$.
}
\label{table:cutflow4}
{\footnotesize
\begin{minipage}[b]{0.45\linewidth}\centering
  \begin{tabular}{c|*4c}
 \toprule
&\multicolumn{4}{c}{ Through $Z$ boson }\\
 \midrule
  $m_{\widetilde{\nu}_{\tau}}$ (GeV)&50&60&80&100\\  \hline
  $p_{\rm T_{\text{trigger}}}$&5,344&6,386&7,458&8,149\\  
  dilepton sel. &4,312&5,203&6,138&6,718\\  
   \bottomrule
  $\epsilon_{\text{sel}}$&0.4312&0.5203&0.6138&0.6718\\  \hline\hline
 \end{tabular}
 \end{minipage}
\hspace{-0.45cm}
\begin{minipage}[b]{0.45\linewidth}
\centering
  \begin{tabular}{|*4c}
 \toprule
 \multicolumn{4}{|c}{Through $W$ boson}\\
 \midrule
  50&60&80&100\\  \hline
  4,929&5,751&6,961&7,698\\  
  3,901&4,652&5,647&6,226\\ 
   \bottomrule
  0.3901&0.4652&0.5647&0.6226\\  \hline\hline
 \end{tabular}
\end{minipage}
\begin{center}
\footnotesize
\begin{minipage}[b]{0.45\linewidth}\centering
  \begin{tabular}{c|*4c}
 \toprule
&\multicolumn{4}{c}{ Through $\gamma ,Z$ bosons}\\
 \midrule
&50&60&80&100\\  \hline
&6,836&5,971&6,912&7,698\\  
&3,089&4,883&5,629&6,257\\  
   \bottomrule
&0.3089&0.4883&0.5629&0.6257\\  \hline\hline
 \end{tabular}
 \end{minipage}
 \end{center}
}
\end{table}

\begin{table}[t]
 \caption{The same as in Table~\ref{table:cutflow3} but for
$\widetilde{\nu}_{\tau}\to ee$.
}
\label{table:cutflow5}
{\footnotesize
\begin{minipage}[b]{0.45\linewidth}\centering
  \begin{tabular}{c|*4c}
 \toprule
&\multicolumn{4}{c}{ Through $Z$ boson }\\
 \midrule
  $m_{\widetilde{\nu}_{\tau}}$ (GeV)&50&60&80&100\\  \hline
  $p_{\rm T_{\text{trigger}}}$&4,724&5,766&6,892&7,633\\  
  dilepton sel. &1,886&2,413&3,011&3,428\\
   \bottomrule
  $\epsilon_{\text{sel}}$&0.1886&0.2413&0.3011&0.3428\\  \hline\hline
 \end{tabular}
 \end{minipage}
\hspace{-0.45cm}
\begin{minipage}[b]{0.45\linewidth}
\centering
  \begin{tabular}{|*4c}
 \toprule
 \multicolumn{4}{|c}{Through $W$ boson}\\
 \midrule
  50&60&80&100\\  \hline
  4,177&4,943&6,046&6,750\\  
  1,627&1,996&2,476&2,797\\  
   \bottomrule
  0.1627&0.1996&0.2476&0.2797\\  \hline\hline
 \end{tabular}
\end{minipage}
\begin{center}
\footnotesize
\begin{minipage}[b]{0.45\linewidth}\centering
  \begin{tabular}{c|*4c}
 \toprule
&\multicolumn{4}{c}{ Through $\gamma ,Z$ bosons }\\
 \midrule
&50&60&80&100\\  \hline
&4,414&5,132&6,078&6,836\\  
&1,868&2,200&2,739&3,089\\  
   \bottomrule
&0.1868&0.2200&0.2739&0.3089\\  \hline\hline
 \end{tabular}
 \end{minipage}
 \end{center}
}
\end{table}

After all, we use the following criteria for the optimized 8-TeV analysis: 
\begin{itemize}
 \item[$\bullet$] At least one muon with $p_{\rm T}>24$~GeV.
 \item[$\bullet$] One pair $\mu^+\mu^-$ or $e^\pm\mu^\mp$ with $p_{\rm
	      T} >10$~GeV and $0.02<|\eta|<2.5$ for each one.
\end{itemize}
The effect of these selections for different masses and for the three
production processes,
is shown in
Tables~\ref{table:cutflow1} and~\ref{table:cutflow2} for the $\mu\mu$
and $\mu e$ channels, respectively. We see that a sizable number of
signal events is expected to pass the selection criteria.
We can compare these results with those of Table~\ref{new}.
For instance, there
$\epsilon_{\rm sel} \simeq 0.11$ is obtained for a 50-GeV sneutrino produced via a
$Z$-boson, whereas $\epsilon_{\rm sel} \simeq 0.66$ is obtained in
Table~\ref{table:cutflow1}, with a significant improvement in the
event selection.

We also study the prospects for the 13-TeV LHC run. Since we do not have
any dedicated searches for dilepton displaced vertices with the 13-TeV
data so far, we just assume background-free in our estimation. Again, we
consider an optimization of the trigger requirements in the 13-TeV
analysis using the existing result for the performance of the ATLAS
trigger system \cite{Aaboud:2016leb}, taking account of the trigger
rate for the 8-TeV analysis. Since the trigger rate for {\tt mu24i} is
$\lesssim 100$~Hz \cite{Aad:2014sca}, we expect a sufficiently
low background as long as the trigger rate in the 13-TeV searches
does not exceed about $100$~Hz. According to Ref.~\cite{Aaboud:2016leb}, a
$p_{\rm T}$ threshold of 26~GeV \cite{ATLAS:muontrigger3,
ATLAS:electrontrigger2} for both muon and electron ensures the trigger
rate to be $\lesssim 100$~Hz.\footnote{Given a higher instantaneous
luminosity ($\sim 2\times 10^{34}~{\rm cm}^{-2}{\rm s}^{-1}$) compared
with those considered in Ref.~\cite{Aaboud:2016leb}, the momentum
threshold is raised from the ones in Ref.~\cite{Aaboud:2016leb} so that
the trigger rates are kept at a similar level.} 
These triggers again rely on the use of the inner tracker,\footnote{For
the tracking performance in the 13~TeV run, see
Ref.~\cite{ATL-PHYS-PUB-2015-018}. } and thus are
effective in the region of $|\eta|<2.5$.
With this observation, we use the
following criteria for the 13-TeV analysis:
\begin{itemize}
 \item[$\bullet$] At least one electron or muon with $p_{\rm T}>26$~GeV.
 \item[$\bullet$] One pair $\mu^+\mu^-$, $e^+e^-$, or $e^\pm\mu^\mp$ with $p_{\rm
	      T} >10$~GeV and $0.02<|\eta|<2.5$ for each one.
\end{itemize}
The effect of these selections for different masses and for the three
production processes at $\sqrt{s}=13$~TeV,
is shown in
Tables~\ref{table:cutflow3}, \ref{table:cutflow4},
and~\ref{table:cutflow5} for the $\mu\mu$, $\mu e$, and $ee$ channels,
respectively.\footnote{The efficiency of the $ee$ channel is 
worse than the other channels due to the isolation requirement for
electrons implemented in the detector simulation with
\texttt{DELPHES v3} \cite{deFavereau:2013fsa}.} We however note that a more elaborate optimization may be
considered; for example, we may also use the dilepton triggers, which
may be more effective since we can lower the momentum threshold for
these triggers \cite{Aaboud:2016leb}. In any case, to use a High Level
Trigger with inner-detector information is technically quite
feasible and expected to result in a considerable improvement in
displaced-vertex searches. We also note in passing that this possible
improvement is not only for the ATLAS analysis but also for the CMS one
\cite{CMS:2014hka}, where again tracker information is not used in
the trigger requirement.

Now we discuss how to obtain the limits for light sneutrinos. 
The limits from the ATLAS search \cite{Aad:2015rba}
can be translated into a vertex-level efficiency, taking into account
the lack of observation of events for any value of the decay
length. Therefore, $\epsilon_{\text{vert}}(c\tau)$ can be obtained as
the ratio of the number of signal events compatible with zero observed
events {(which in this case is 3)} and that corresponding to the upper limits given in
Ref.~\cite{Aad:2015rba} (with an appropriate modification described
above); for example, we can use
{the purple-shaded solid line of} Fig.~\ref{Fig13} to obtain the vertex-level efficiency
$\epsilon^{\mu\mu}_{\text{vert}}(c\tau)$ for the dimuon channel. By multiplying the
number of the events passing the trigger and event selection criteria, which
is computed above, with this vertex-level efficiency, we can estimate
the total number of signal events; for the 8-TeV case, {this is given 
for the $\mu\mu$ channel 
by
\begin{eqnarray}
{\#} \text{Dimuons} &=&
\left[\sigma(pp\to Z\to\widetilde{\nu}_{\tau}\widetilde{\nu}_{\tau})
\epsilon_{\text{sel}}^Z
+
\sigma(pp\to W\to\widetilde{\nu}_{\tau}\widetilde{\tau})
\epsilon_{\rm sel}^W
+\sigma(pp\to \gamma,Z\to\widetilde{\tau}\widetilde{\tau})
\epsilon_{\text{sel}}^{\gamma ,Z}
\right] 
\nonumber\\
&\times& 
\mathcal{L}\times \left[\text{BR}(\widetilde{\nu}_{\tau}^\mathcal{R}\to\mu\mu)\
\epsilon^{\mu\mu}_{\text{vert}}(c\tau^\mathcal{R})+\text{BR}
(\widetilde{\nu}_{\tau}^\mathcal{I}\to\mu\mu)\
\epsilon^{\mu\mu}_{\text{vert}}(c\tau^\mathcal{I})\right],
 \label{numberevents}
\end{eqnarray}
where the selection efficiencies $\epsilon_{\rm sel}^Z$, $\epsilon_{\rm
sel}^W$ and $\epsilon_{\rm sel}^{\gamma ,Z}$ are given {in 
Tables~\ref{new} and~\ref{table:cutflow1}.
The same formula can be applied for the $e\mu$ channel shown in 
Tables~\ref{new2} and~\ref{table:cutflow2}, using the corresponding BRs, selection efficiencies, and vertex-level efficiencies (which turn out to be similar).}
For the 13-TeV
prospects the selection efficiencies {for the three channels} can be found in 
Tables~\ref{table:cutflow3}--\ref{table:cutflow5}, and we use the same vertex-level efficiency {as in the 8-TeV case} and assume zero
background}.\footnote{Notice that in the 13-TeV long-lived gluino search
\cite{Aaboud:2017iio} the estimated number of background events is still
much smaller than {one}, $\sim 10^{-2}$, which is similar in size to that
in the 8-TeV search \cite{Aad:2015rba}. We therefore expect that the
background in the 13-TeV dilepton displaced-vertex search is also as low
as the 8-TeV one.} As a
result, if this predicted number of signal {events is above 3}
the corresponding parameter
point of the model is excluded so that this is compatible with zero
number of events. 

We note in passing that the optimization strategy we have discussed in
this section is generically useful for the searches of other metastable
particles which have a relatively short lifetime and a small mass and
whose decay products contain soft leptons. We thus hope this kind
of search strategies to be considered seriously in the LHC experiments. 

Another possibility to improve the sensitivity is to search for
shorter displaced vertices. As can be seen from Fig.~\ref{fig:BRb}, in
some regions of parameter space, the decay distance of a sneutrino is
predicted to be $\lesssim 1$~mm, to which the current ATLAS search is
less sensitive. However, we may even probe such sub-millimeter region by
relaxing the impact-parameter requirements for lepton tracks as well as
the condition on the reconstructed position of displaced vertices, given the extremely
low background in this search \cite{Aad:2015rba}. In fact, it is shown
in Refs.~\cite{Ito:2017dpm, Ito:2018asa} for metastable gluinos that sub-millimeter
displaced vertices can be probed using the existing
vertex-reconstruction technique, though this is not directly applicable
to the present case. Moreover, there are several existing
searches which are sensitive to sub-millimeter region
\cite{CMS:2014hka, Khachatryan:2014mea, Khachatryan:2016unx, CMS-PAS-EXO-16-022}. In
any case, to assess the possibility of searching for shorter dilepton
displaced vertices, a dedicated study with a full consideration of the detector
performance is required; we thus do not discuss this possibility in this
paper and hand this over to experimentalists.

\section{Results}
\label{results}

By using the method described in the previous section, we now evaluate the current and
potential limits on the 
parameter space of our scenario from the displaced-vertex
searches with the 8-TeV ATLAS result \cite{Aad:2015rba}, and discuss the
prospects for the 13-TeV searches.


\begin{figure}[t!]
\centering
\includegraphics[scale=0.5]{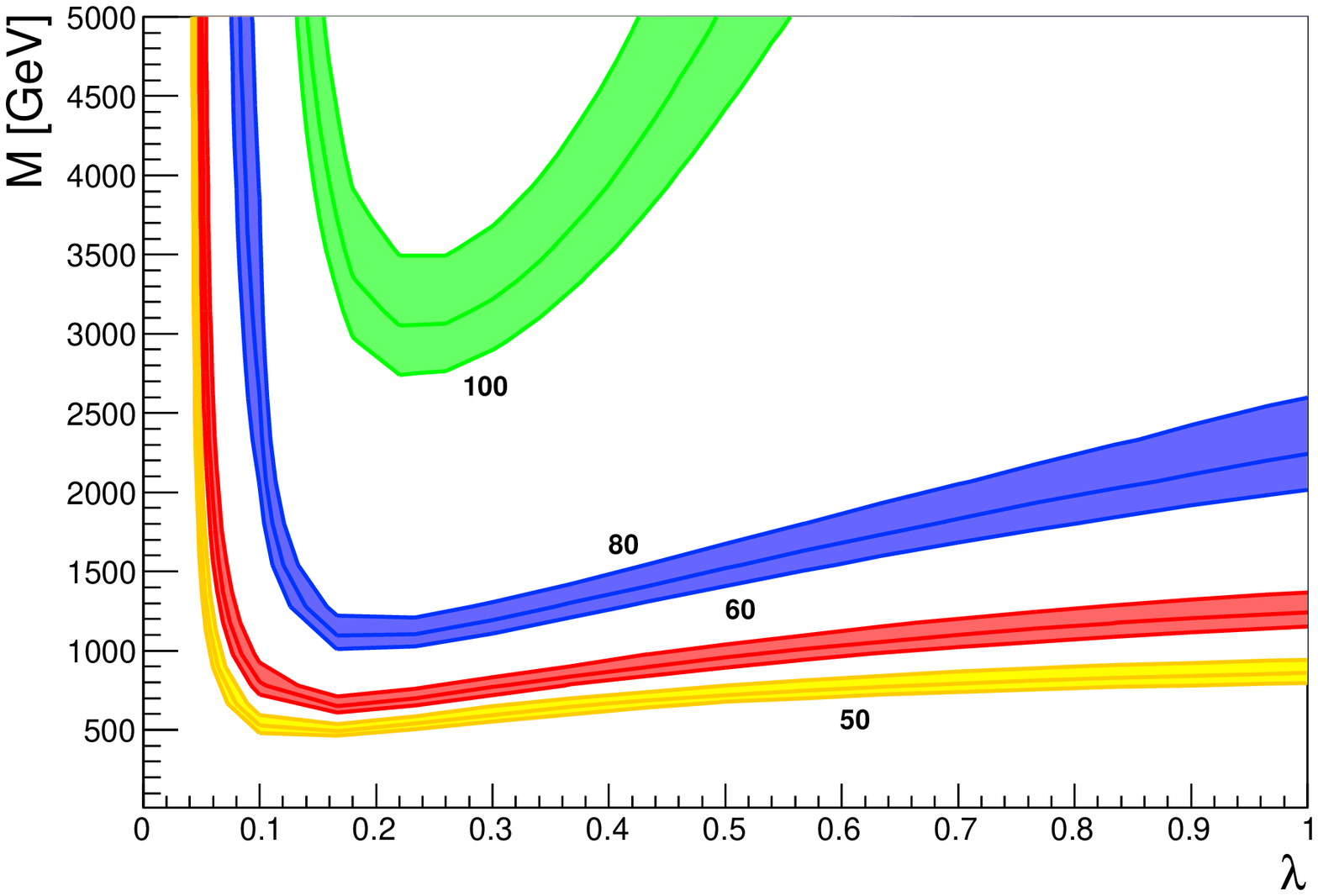}
\includegraphics[scale=0.5]{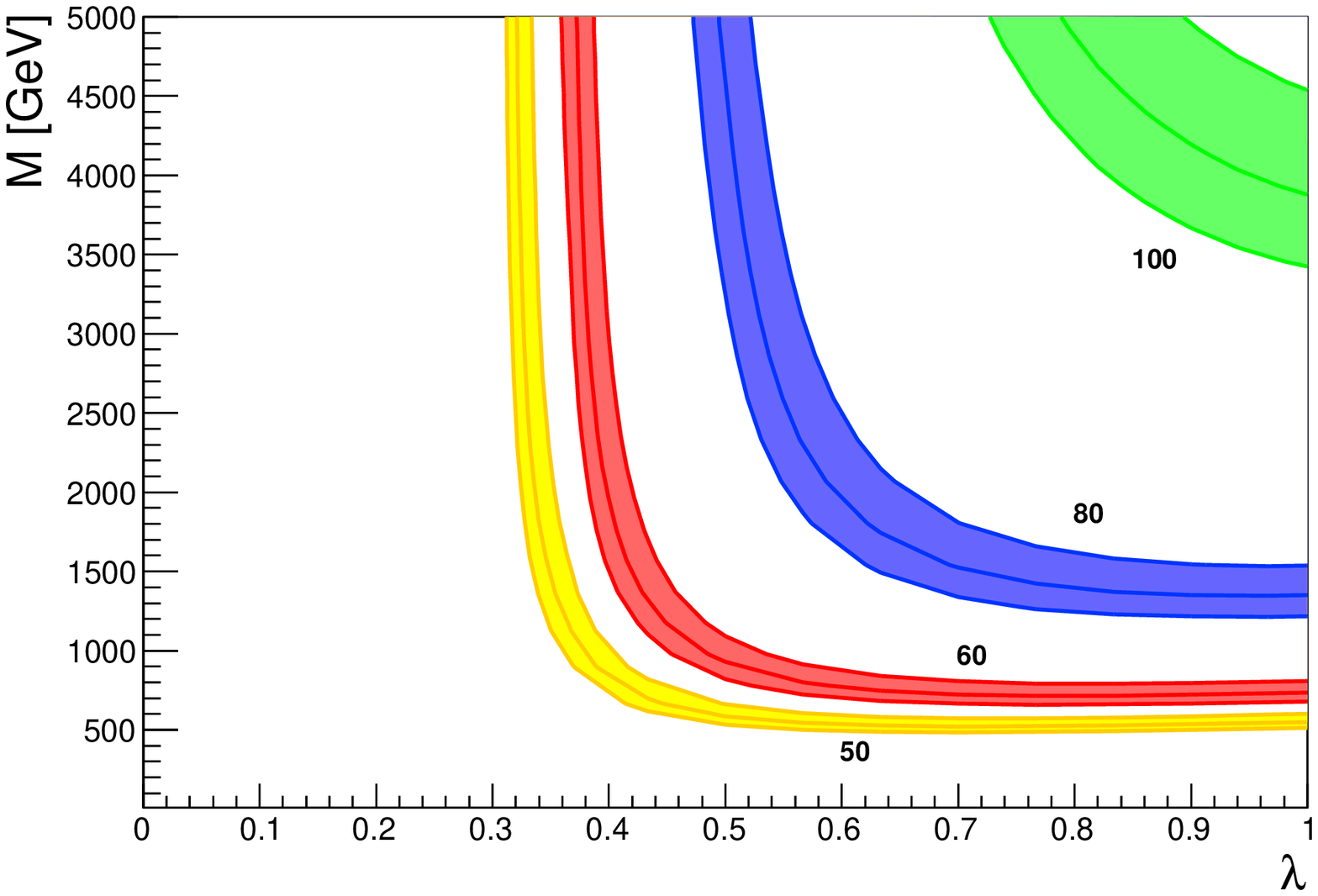}
\includegraphics[scale=0.5]{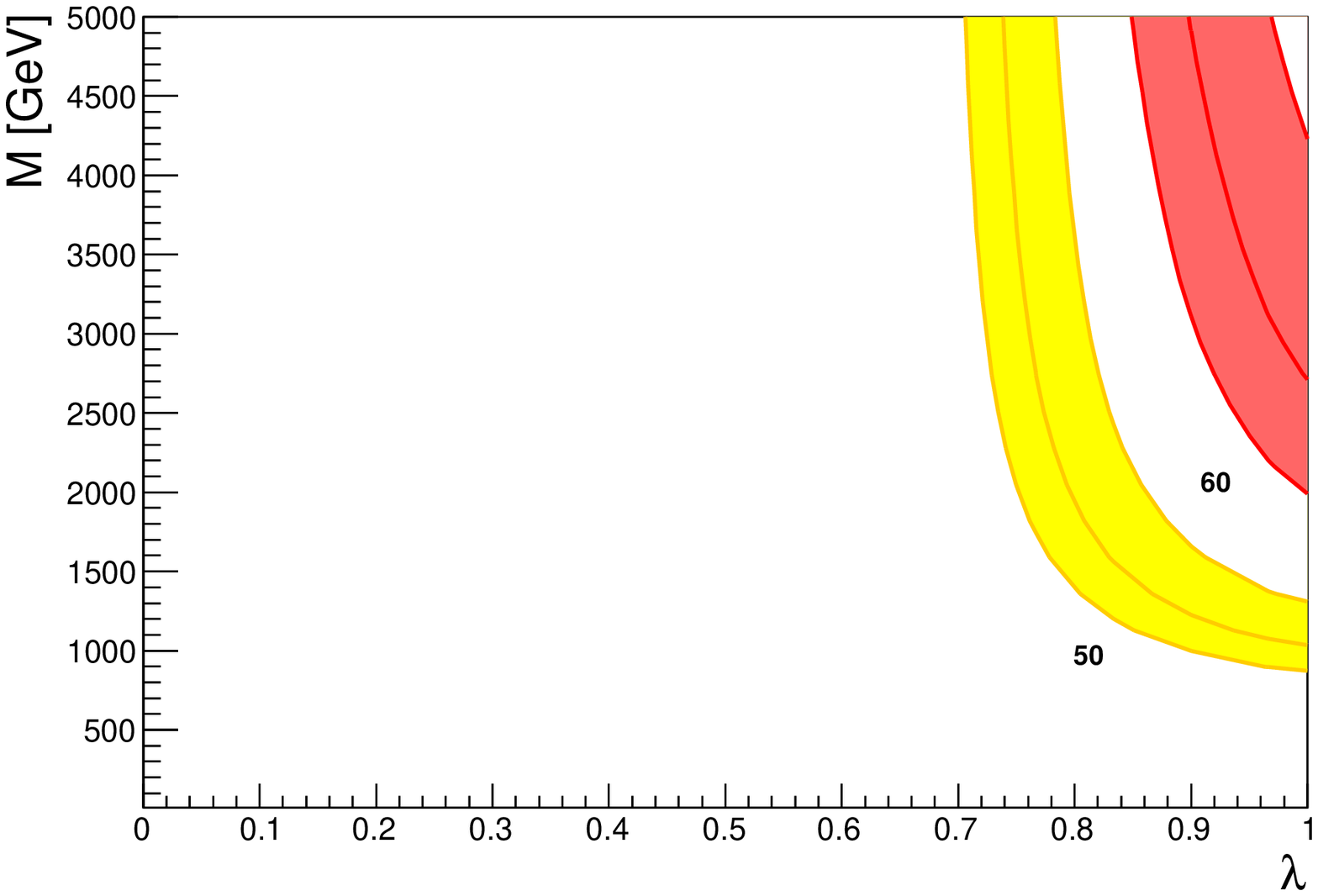}
\caption{Limits on the 
parameter space of our scenario from the ATLAS 8-TeV displaced-vertex search with an integrated luminosity of 20.3~fb$^{-1}$~\cite{Aad:2015rba}, combining the $\mu\mu$ and $e\mu$ channels. The region {to the right and above of} each colored line is excluded.
The neutrino mass scale is fixed to be $m_{\nu}\sim 0.05$~eV, and the
neutrino Yukawa couplings are set to $Y_\nu =10^{-7}$, $5\times
10^{-7}$, and $10^{-6}$ in the top, middle, and bottom
panels, respectively.
The yellow, red, blue, and green lines correspond to the sneutrino mass of 50, 60, 80 and 100 GeV, respectively, {with the
bands representing the uncertainties that come from those in Fig.~13 (c) of
Ref.~\cite{Aad:2015rba}}. 
}
\label{fig:limits0.05non}
\end{figure}

 
 \begin{figure}[t!]
\centering
\includegraphics[scale=0.5]{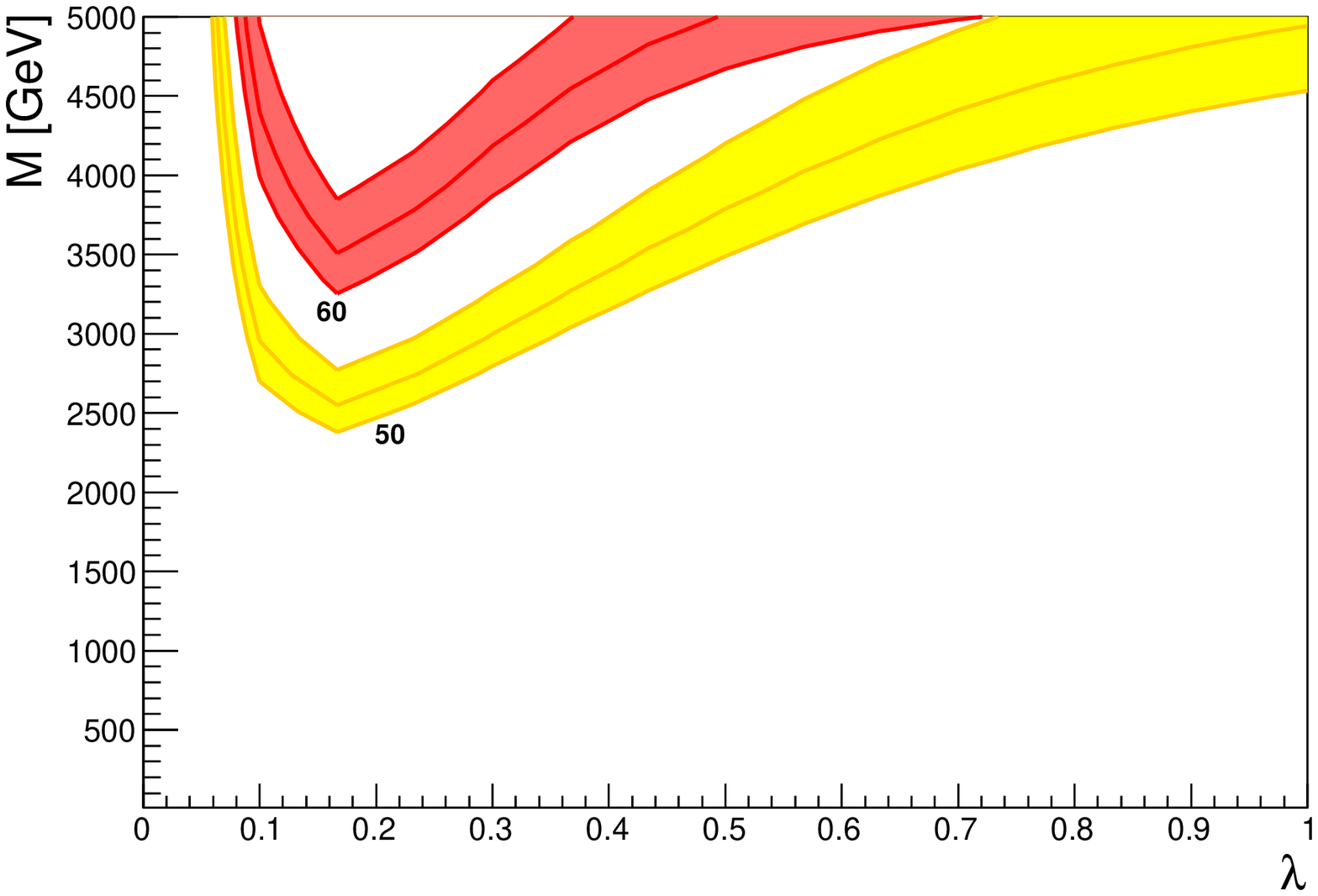} 
\includegraphics[scale=0.5]{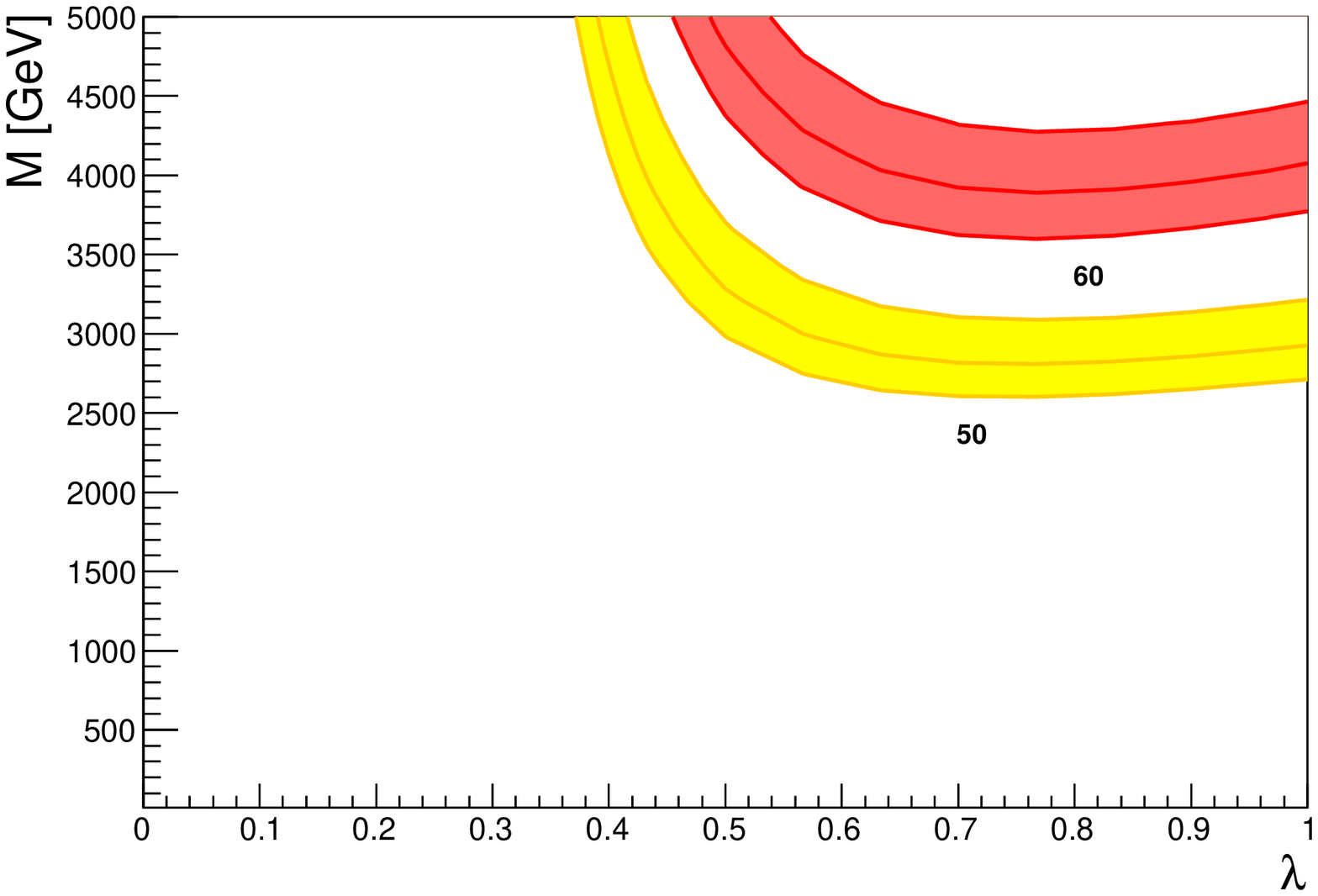}
\includegraphics[scale=0.5]{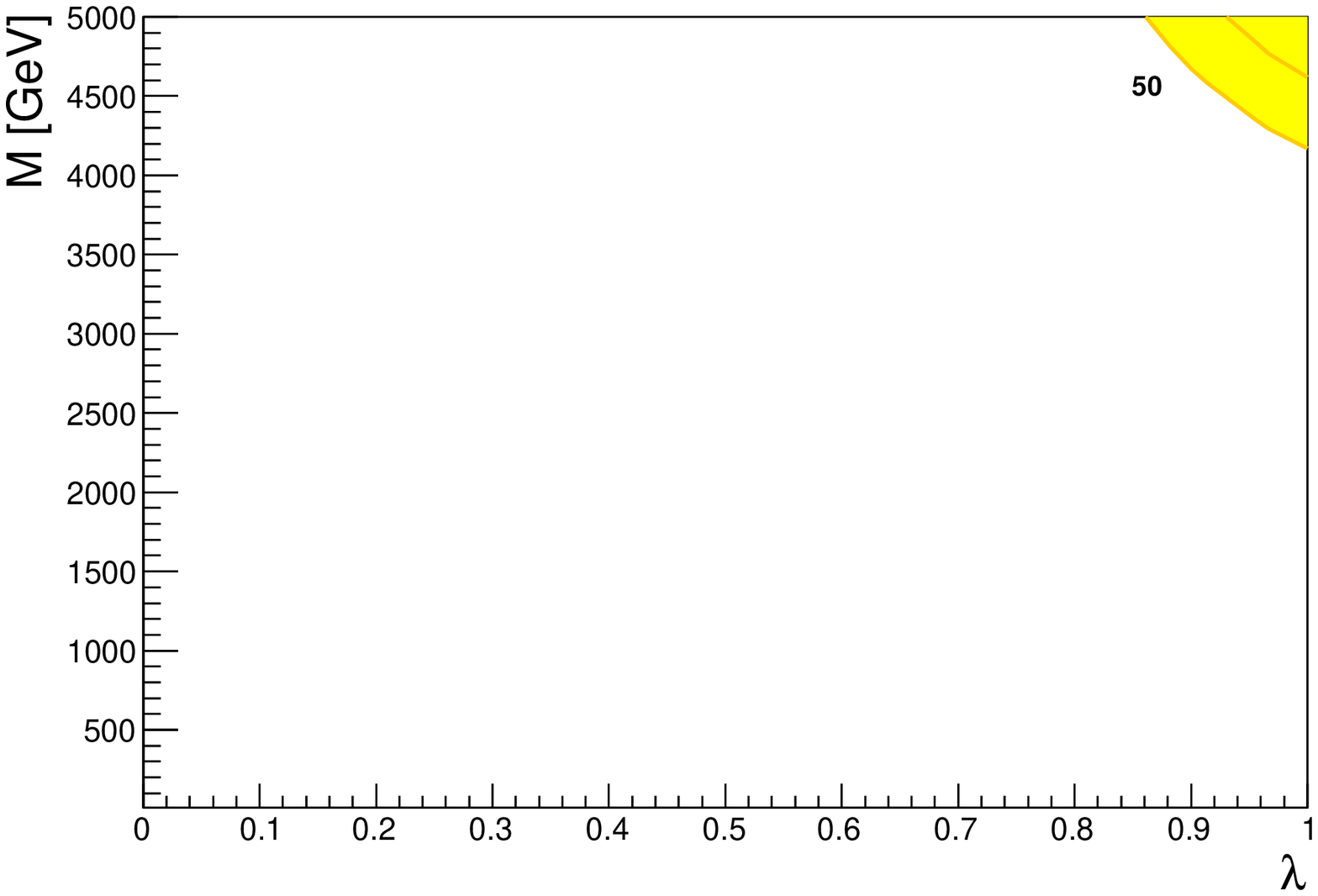} 
\caption{The same as in Fig.~\ref{fig:limits0.05non} but with the neutrino mass scale fixed to $m_{\nu}\sim 0.23$~eV.}
\label{fig:limits0.23non}
\end{figure}

\begin{figure}[t!]
\centering
\includegraphics[scale=0.55]{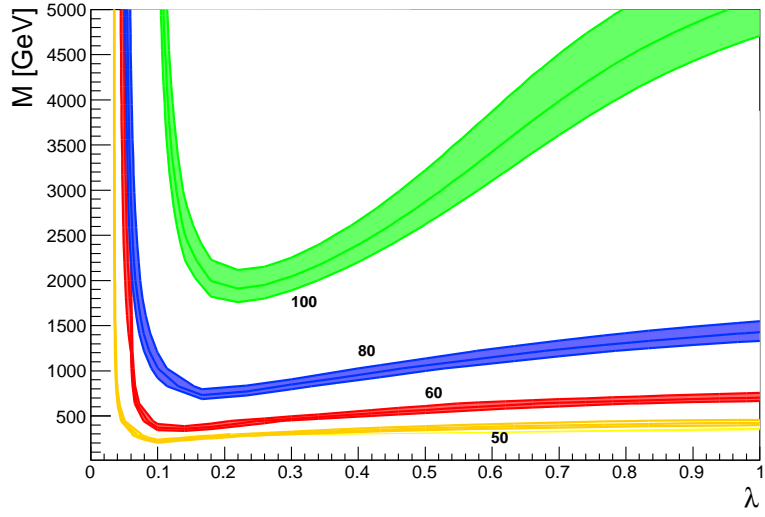} 
\includegraphics[scale=0.55]{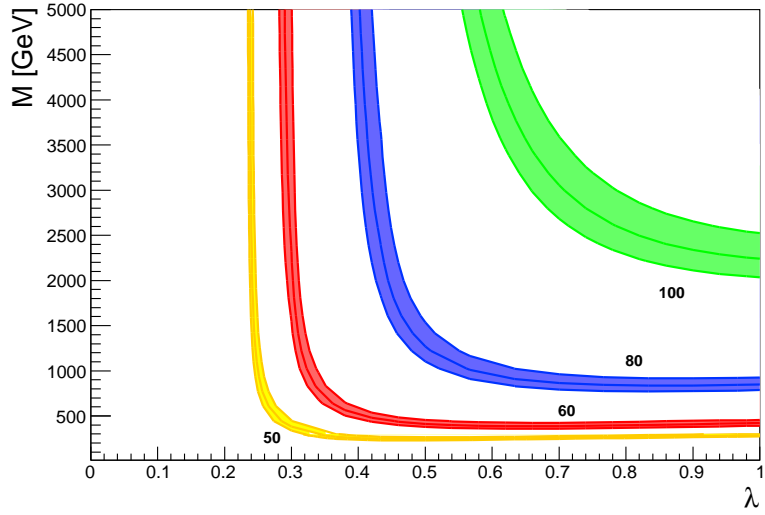}
\includegraphics[scale=0.55]{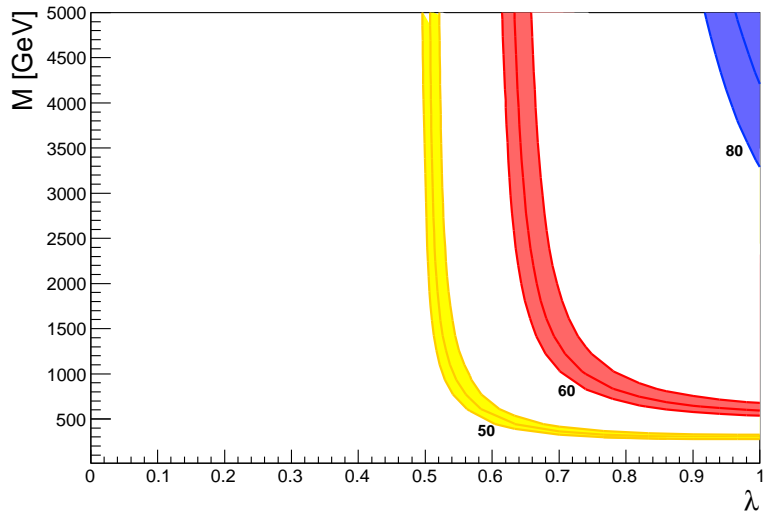}
\caption{
The same as in Fig.~\ref{fig:limits0.05non} where $m_{\nu}\sim 0.05$~eV,
but considering the optimization of the trigger requirements discussed in the text.
}
\label{fig:limits0.05}
\end{figure} 

\begin{figure}[t!]
\centering
\includegraphics[scale=0.55]{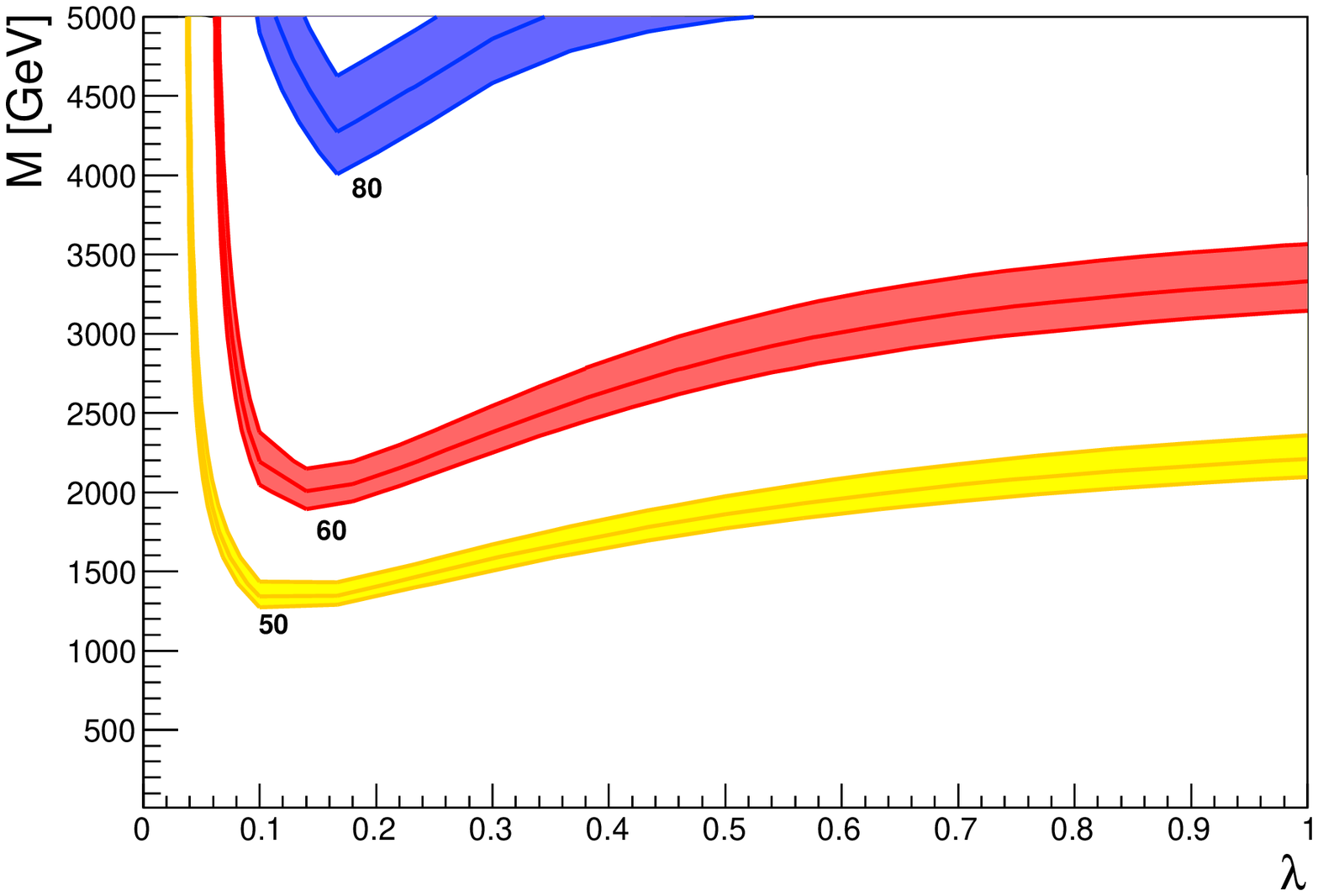} 
\includegraphics[scale=0.55]{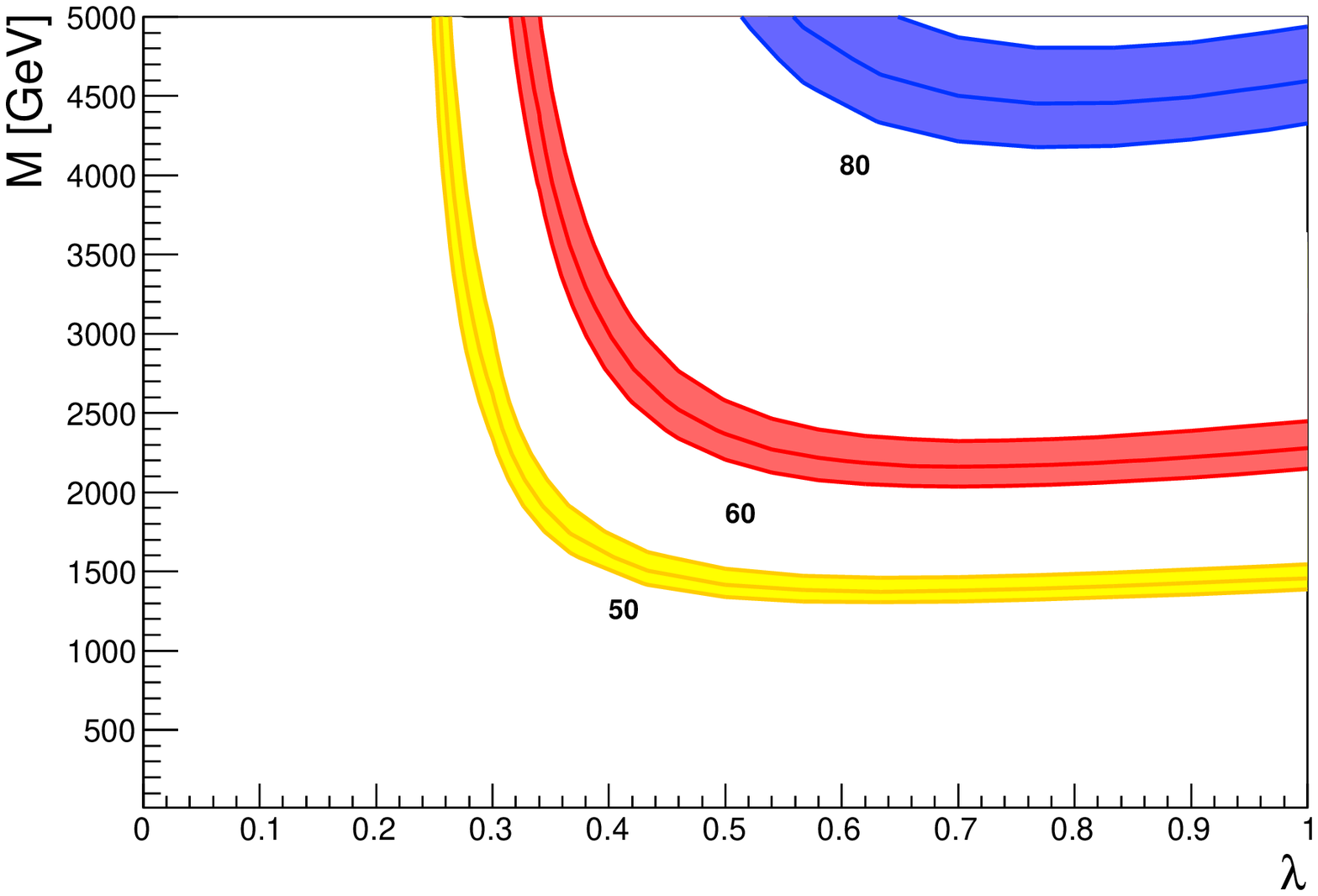}
\includegraphics[scale=0.55]{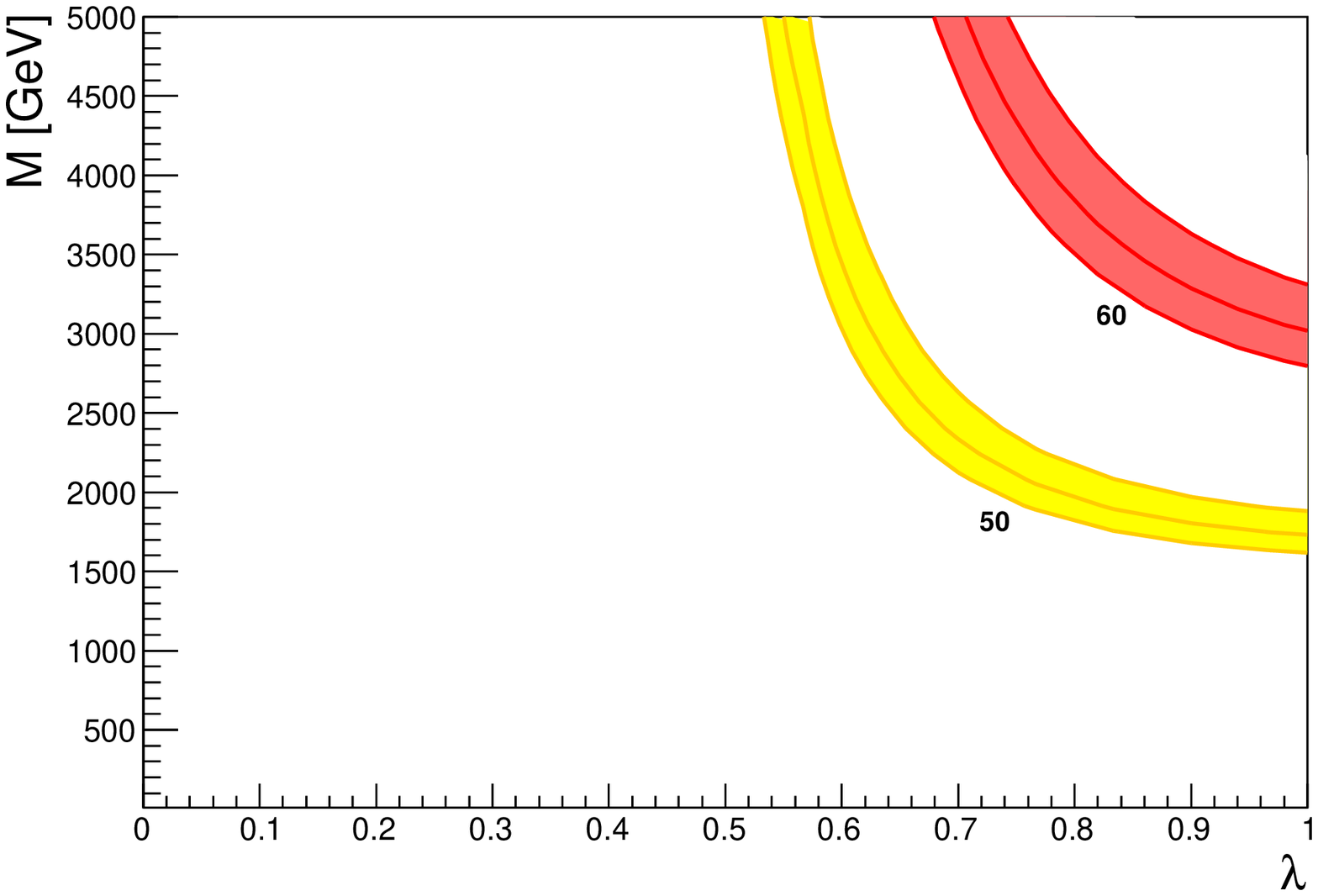}
\caption{The same as in Fig.~\ref{fig:limits0.05} but with the neutrino mass scale fixed to $m_{\nu}\sim 0.23$~eV.}
\label{fig:limits0.23}
\end{figure} 



The 8-TeV current limits are given in 
Figs.~\ref{fig:limits0.05non} and~\ref{fig:limits0.23non}, with
the neutrino mass scale fixed to be 0.05 and 0.23 eV, respectively, and 
all of the neutrino Yukawa
couplings set to be a common value $Y_\nu$: $10^{-7}$, $5\times
10^{-7}$, and $10^{-6}$~GeV in the top, middle, and bottom
panels, respectively. 
The yellow, red, blue, and green lines correspond
to the sneutrino mass of 50, 60, 80, and 100~GeV, respectively, with the
bands representing the uncertainties that come from those in
Ref.~\cite{Aad:2015rba}, {i.e. the $\pm 1\sigma$ variations in the expected limit in Fig.13 (c) of the aforementioned reference}.
Notice that for these sneutrino masses, values of $M$ smaller than 92.7, 111.3, 148.4, and 185.5 GeV, respectively, are not interesting for our analysis since the tau left sneutrino would no longer be the LSP in favor of the gauginos.
To obtain the reaches, we have combined the
results from the $\mu\mu$ and $e\mu$ channels.
The region of the parameter space {to the right and above of} each line is excluded from the displaced-vertex searches.

We can see in the top panel of Fig.~\ref{fig:limits0.05non}, where
$Y_\nu = 10^{-7}$, that for $m_{\widetilde{\nu}_{\tau}}=50$ GeV the
upper bound on the average gaugino mass $M$ (see Eq.~(\ref{eq:3.550}))
is of about 500 (900) GeV for $\lambda = 0.1$ (1). 
As discussed in Section~\ref{phenomenology}, small values of $\lambda$
favor larger BRs, and as a consequence the gaugino mass is more
constrained. If $\lambda$ is too small, however, the limit on $M$
disappears since the lifetime of the left sneutrino goes into the
sub-millimeter regime, as can be seen from Eq.~\eqref{eq:3.22112}. 
On the other hand, small sneutrino masses produce larger decay lengths,
and gaugino masses turn out to be also more constrained. For example,
for  $m_{\widetilde{\nu}_{\tau}}=50$ and $60$~GeV the upper bound on $M$ 
for $\lambda =1$ is
of about 700 and 1100 GeV, respectively.
In the middle panel, the larger value of the neutrino Yukawa $Y_\nu = 5\times 10^{-7}$ gives rise to smaller decay lengths, and therefore the figures are shifted to the right with a lower limit for $\lambda$ of about 0.3.
The case of $Y_\nu = 10^{-6}$ in the bottom panel is more extreme, and the lower limit on 
 $\lambda$ is now of about 0.7.
 Finally, in Fig.~\ref{fig:limits0.23non} we show the same cases as in 
 Fig.~\ref{fig:limits0.05non} but for the neutrino mass scale 0.23 eV.
To increase the neutrino mass produces an increase in the left sneutrino VEVs, and therefore the $\widetilde{\nu}_{\tau}$ decay width into neutrinos is larger.
As a consequence, its BR into leptons as well as its decay length are
smaller giving rise to less stringent constraints on the parameter
space, as shown in the figure. In particular, there is almost no
constraint on the parameter space shown in the bottom panel in
Fig.~\ref{fig:limits0.23non}.

The potential limits using the optimization of the trigger requirements explained in the previous section, are shown in 
Figs.~\ref{fig:limits0.05}  and~\ref{fig:limits0.23}. They turn out to be more stringent than the previous ones without optimization.
{For example, we can see in the top panel of Fig.~\ref{fig:limits0.05}  that for $m_{\widetilde{\nu}_{\tau}}=50$ GeV the upper bound on $M$ is of about 200 (500) GeV for
$\lambda =$  0.1 (1), to be compared with the ones of
Fig.~\ref{fig:limits0.05non}. Moreover, in the case of $Y_\nu=10^{-6}$ and
$m_\nu \sim 0.23$~eV shown in the bottom panel in
Fig.~\ref{fig:limits0.23}, now some limits are imposed on light sneutrino
masses, which can be compared to the bottom panel in
Fig.~\ref{fig:limits0.23non}. This result indicates that the trigger
optimization discussed above can significantly improve the sensitivity of
left sneutrino searches.  

%
\begin{figure}[t!]
\centering
\includegraphics[scale=0.50]{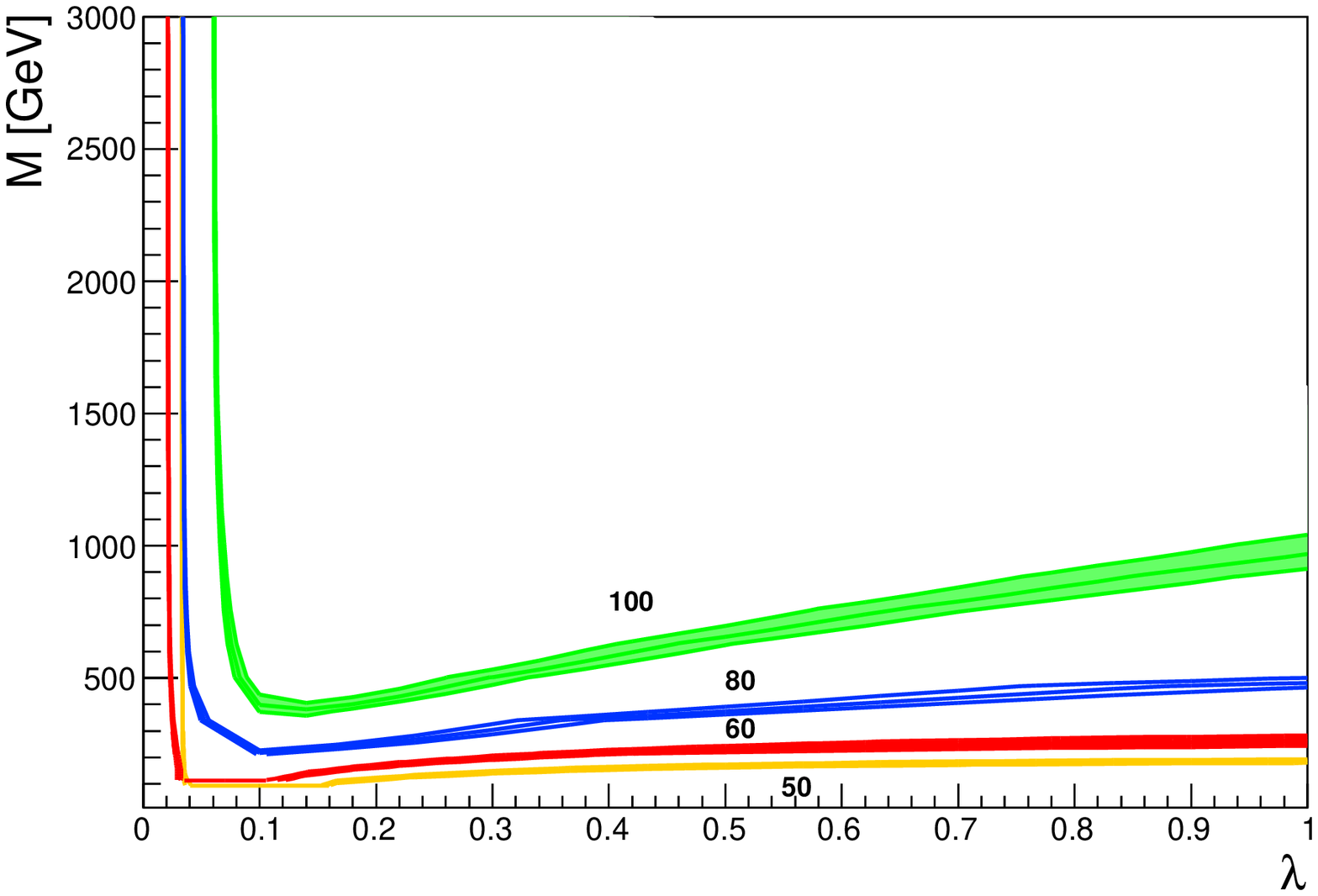} 
\includegraphics[scale=0.50]{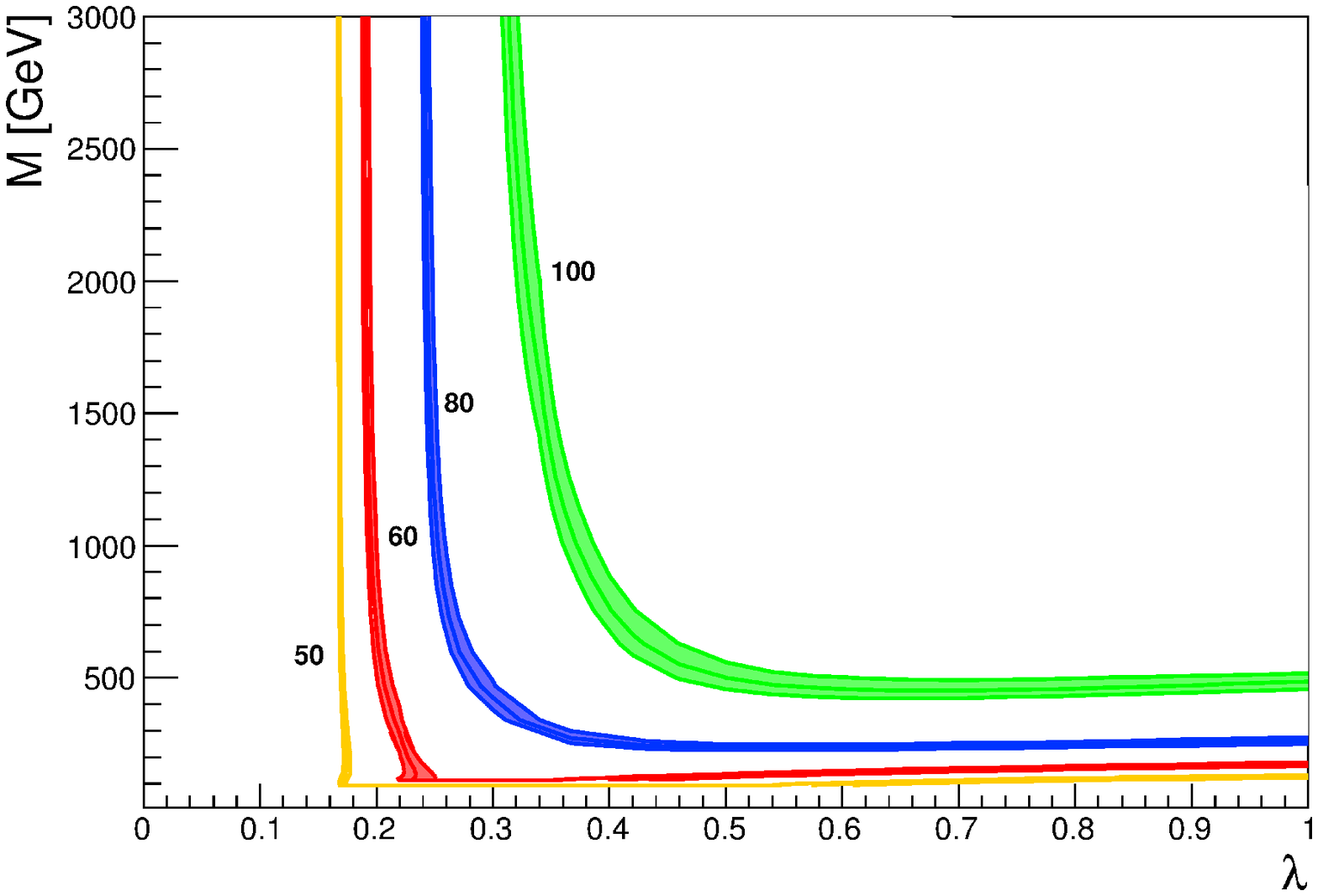}
\includegraphics[scale=0.50]{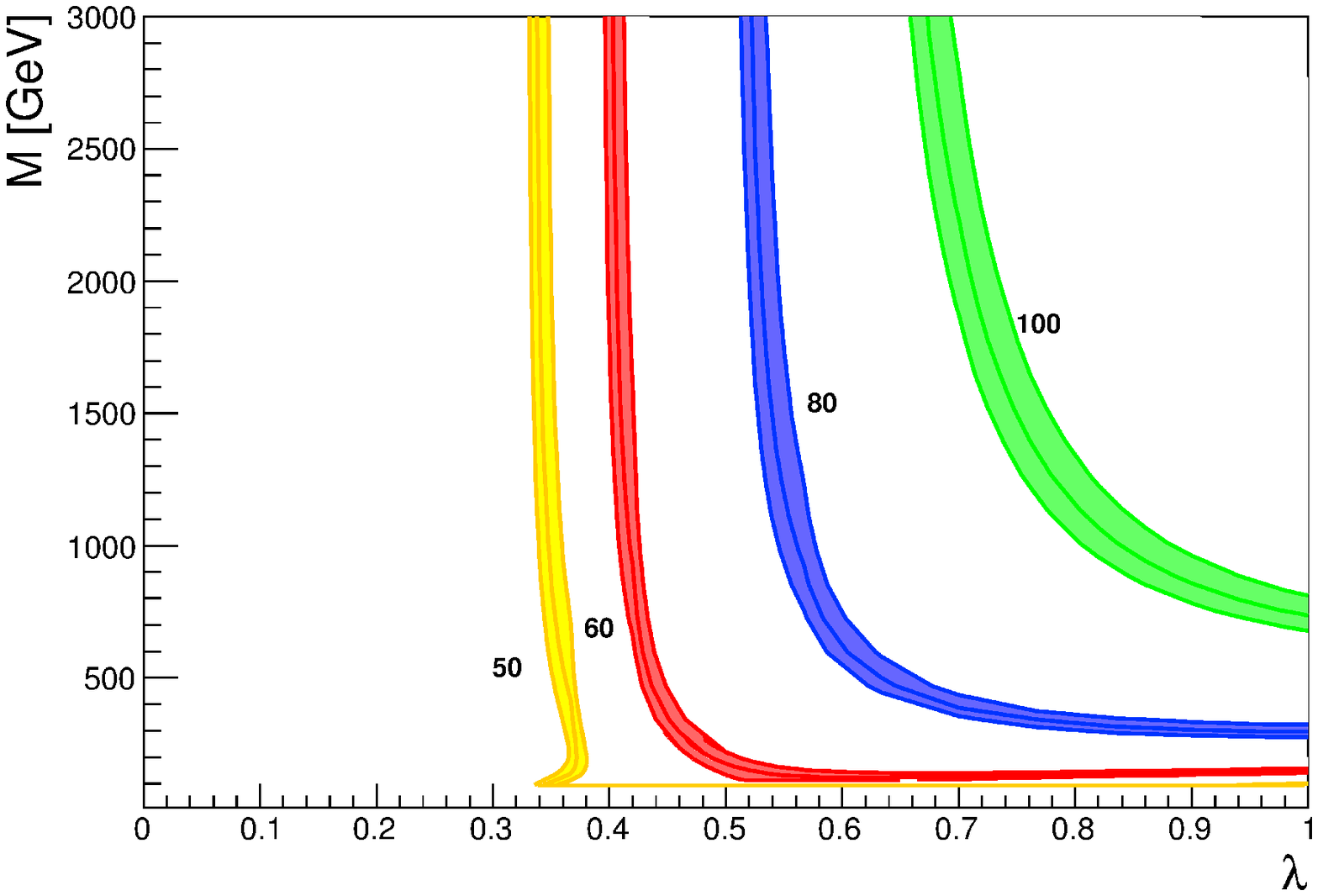}
\caption{
The same as in Fig.~\ref{fig:limits0.05non} where $m_{\nu}\sim 0.05$~eV, but analyzing the prospects for the 13-TeV search with an integrated luminosity of 300~fb$^{-1}$, combining the $\mu\mu$, $e\mu$ and $ee$ channels, and considering also the optimization of the trigger requirements discussed in the text.
}
\label{fig:limits0.05_II}
\end{figure} 
\begin{figure}[t!]
\centering
\includegraphics[scale=0.50]{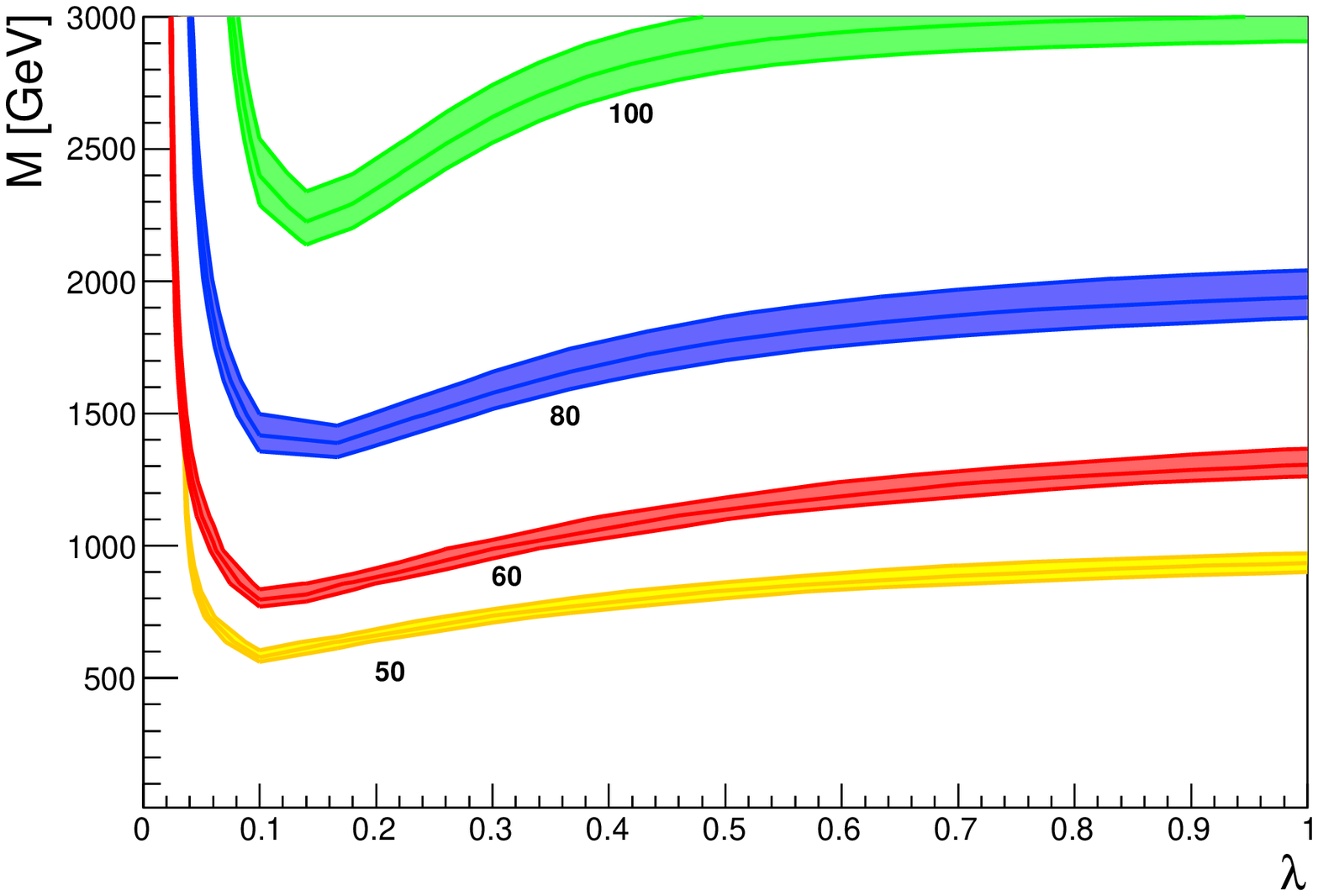} 
\includegraphics[scale=0.50]{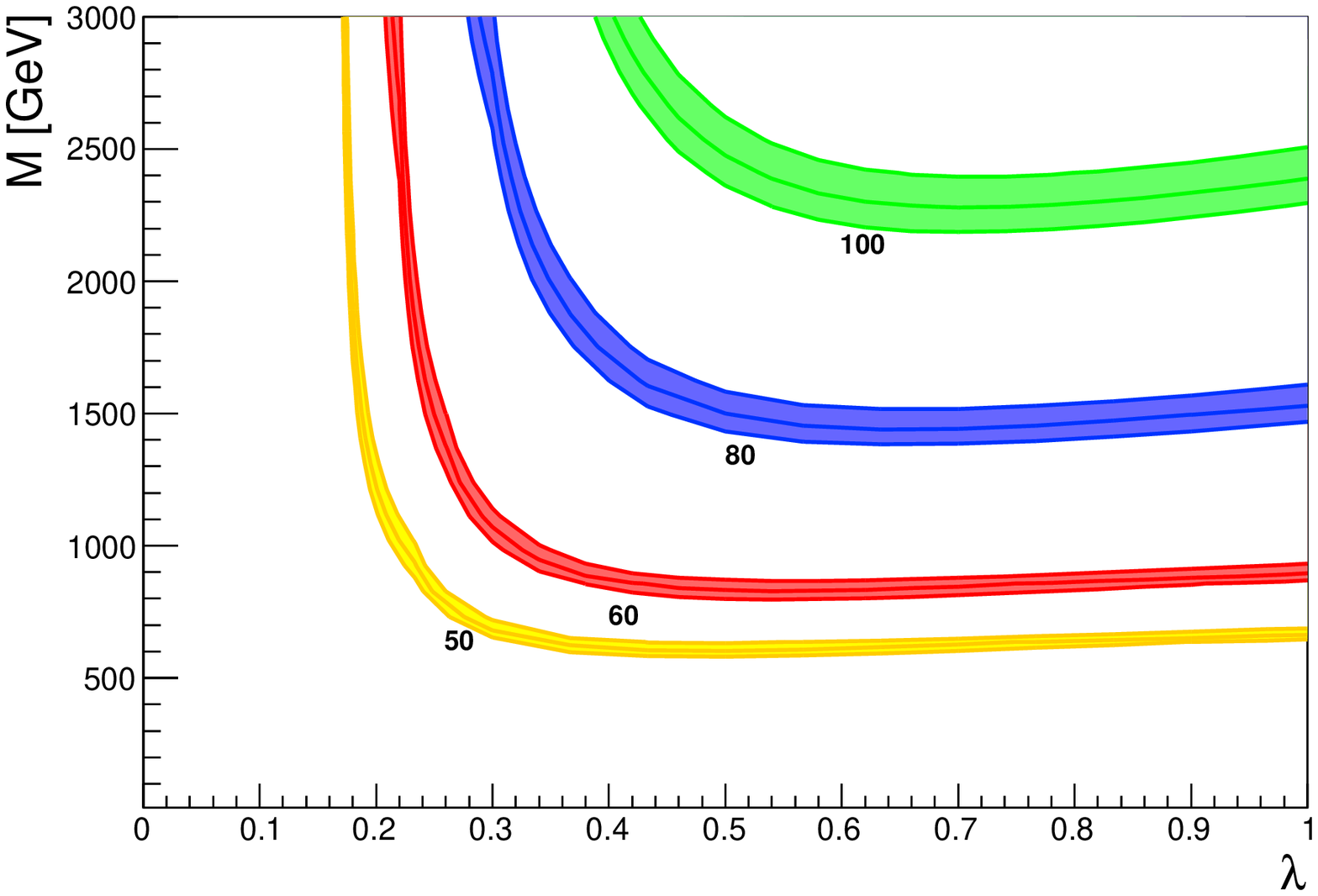}
\includegraphics[scale=0.50]{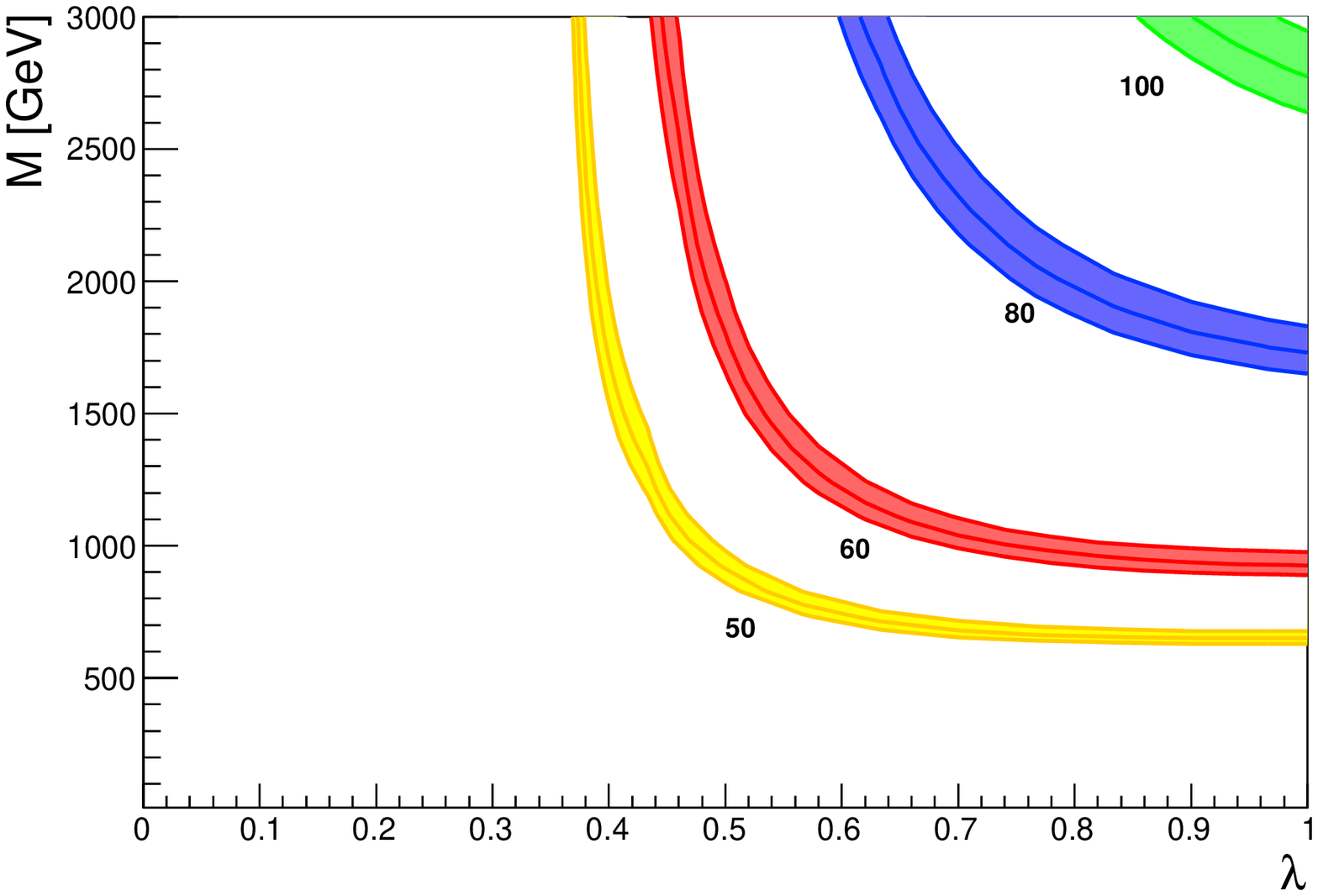}
\caption{
The same as in Fig.~\ref{fig:limits0.05_II} but with the neutrino mass scale fixed to $m_{\nu}\sim 0.23$~eV.
}
\label{fig:limits0.23_II}
\end{figure} 

The 13~TeV prospects are illustrated in Figs.~\ref{fig:limits0.05_II}
and \ref{fig:limits0.23_II}.
Here, we combine the $\mu\mu$, $e\mu$ and $e e$ channels.
As we can see, the constraints on the parameter space of our scenario
turn
out to be very strong. For example, in the top panel of
Fig.~\ref{fig:limits0.05_II} and for $\lambda=0.1$, the upper bound on
$M$ is as small as about 100, 200, and 350 GeV for
$m_{\widetilde{\nu}_{\tau}}\approx 50, 80$, and 100 GeV. 
Such small $M$ region would be probed in other strategies like
ordinary electroweak gaugino searches, which makes it possible to cover a
considerable range of the parameter space for the left sneutrino LSP
with a mass in the range 45--100 GeV. Furthermore, now it is possible to
probe a wide range of the parameter space even for the heavier neutrino mass
case, $m_\nu \sim 0.23$~eV, as shown in Fig~\ref{fig:limits0.23_II}. All in all, we conclude that the dilepton
displaced-vertex searches can be a powerful probe of the $\mu\nu$SSM
parameter space, especially if we optimize them by making the most of
the performance of the inner detectors of the ATLAS and CMS
experiments.

\section{Conclusions and discussions}
\label{conclusions}

We have {analyzed} the sensitivity of the displaced dilepton searches at
the LHC to a tau left sneutrino LSP with a mass in the range
45--100~GeV in the framework of the $\mu\nu$SSM. The sneutrino LSP is
produced via the $Z$-boson mediated Drell-Yan process or through the
$W$- and $\gamma/Z$-mediated process accompanied with the production and
decay of the left stau NLSP. Due to the $R_p$ violating term
present in the $\mu\nu$SSM, the left sneutrino LSP becomes metastable and
eventually decays into the {standard model} leptons. Because of the large
value of the tau Yukawa coupling, a large fraction of the sneutrino LSP
decays into a pair of tau leptons or a tau lepton and a light charged
lepton, while the rest decays into a pair of neutrinos. It is found that
the decay distance of the left sneutrino tends to be as large as $\gtrsim
1$~mm, which thus can be a good target of displaced vertex searches. We
have found that the displaced dilepton search channel is most sensitive to the
sneutrino LSP, where at least one of the pair-produced left sneutrinos
is required to decay into $\tau \tau$ or $\tau \ell$ with the final-state
tau leptons decaying leptonically. To evaluate the prospects of
this search strategy, we recast the result of the ATLAS 8-TeV
dilepton search to obtain the potential limit on the 
parameter space of our scenario from the 8-TeV LHC data. It is found that even the
present data set potentially gives a constraint on the left sneutrino
LSP, especially when the Yukawa couplings and mass scale of neutrinos
are rather small. We have also discussed an
optimization of the trigger requirements exploited in the ATLAS search
based on a High Level Trigger that utilizes the tracker information. It
turns out that this optimization can considerably improve the
sensitivity of the displaced dilepton search. Moreover, we have
estimated the potential limits obtained at the 13-TeV LHC run and found
that wide range of the 
parameter space of our scenario can be probed at the
LHC Run~3.

As mentioned above, we may consider further optimization for the
sneutrino LSP search. Given the low background in the dilepton
displaced-vertex search, we may relax the condition on the impact
parameter of lepton tracks used for the reconstruction of displaced
vertices, as well as that on the reconstructed position of displaced
vertices. With such a relaxation, it may be possible to detect
sub-millimeter dilepton displaced vertices, which allows us to probe
sneutrinos with a shorter lifetime. A further optimization for the
trigger requirements is another interesting option to improve the
potential of this search. For instance, we may also use the dilepton
triggers, which accommodate a lower momentum threshold. Such
optimizations highly rely on the detector performance and thus a more
dedicated study is required to assess their prospects.

Displaced sneutrino decay signature is useful not only for its discovery
but also for the determination of parameters relevant to the sneutrino
decay properties. For example, measurement of the lifetime of the
sneutrino LSP through the reconstruction of displaced vertices allows us
to constrain the parameters in Eq.~\eqref{eq:3.22112}, such as $Y_\nu$,
$M$, and $\lambda$. In addition, it is in principle possible to
measure the mass of the sneutrino LSP since it can decay into
visible particles such as $\tau \tau$ and $\tau \ell$; by using
hadronically decaying tau leptons, we may fully reconstruct the momenta
of the final-state leptons. Although this may be rather challenging, it
is worth investigating this possibility in the future.

In this paper, we focus on the simplest case of the $\mu\nu$SSM with one
right-handed neutrino superfield. Of course, the metastable left
sneutrino can also appear if we introduce three right-handed neutrinos
so that they give masses to light neutrinos at tree level. In this case,
the sneutrino couplings should be chosen so that the neutrino
oscillation data is reproduced, which may have some implications for the
sneutrino decay properties. Another interesting possibility is to
consider a different LSP, which can be again long-lived due to the small
$R_p$ violation. In particular, a colored LSP such as the stop LSP may be
interesting as its production cross section is quite large at the
LHC. Even in this case, we may still probe it by searching for, e.g.,
multi-track displaced vertices. These subjects will be discussed in
another occasion \cite{prepa}.


\begin{acknowledgments}

NN and CM thank K. A. Olive and his family for their hospitality at their
 home, where the present work was initiated, during `Olivefest:
 Astroparticle Physics Looking Forward'.
The work of IL and CM was supported in part by the Spanish Agencia Estatal de Investigaci\'on 
through the grants 
FPA2015-65929-P (MINECO/FEDER, UE) and IFT Centro de Excelencia Severo Ochoa SEV-2016-0597. The work 
of DL was supported by the Argentinian CONICET, and he also acknowledges the support of the Spanish grant FPA2015-65929-P (MINECO/FEDER, UE). 
The work of NN was supported by the Grant-in-Aid for Scientific Research
 (No.17K14270).
The work of RR was supported 
in part by the grant
FPA2014-57816-P,
and the Program 
SEV-2014-0398 `Centro de Excelencia  Severo Ochoa'. 
IL, CM, DL and RR also acknowledge the support of the MINECO's Consolider-Ingenio  2010 Programme under 
grant MultiDark CSD2009-00064.
 
\end{acknowledgments}

\bibliographystyle{JHEP}
\bibliography{munussm_v8_v2}
\end{document}